\documentclass[11pt, twocolumn]{IEEEtran}
\usepackage[utf8]{inputenc}
\usepackage{color}
\usepackage{amsmath}
\usepackage{amssymb} 
\usepackage{graphicx}
\usepackage{xspace}
\usepackage{bbm} 
\usepackage{color}
\usepackage{cite}
\usepackage{multirow}

\usepackage{mathtools}

\newcommand{\unit}[1]{\ 
    \ifmmode & \left[\textup{#1}\right] \hspace{1cm}
    \else \hfill $\left[\textup{#1}\right]$ \hspace{1cm}
    \fi
}
\newcommand{\vectt}[1]{%
  \if#1\relax\bm{#1}\else\mathbf{#1}\fi
}


\begin{document}

\title{A Very Brief Introduction to Machine Learning With Applications to Communication Systems}

\author{Osvaldo Simeone,~\IEEEmembership{Fellow,~IEEE} 
\thanks{
King's College London, United Kingdom (email: osvaldo.simeone@kcl.ac.uk). This work has received funding from the European  Research  Council (ERC) under the European Union Horizon 2020 research and innovation program (grant agreement 725731).}}

\maketitle
\thispagestyle{plain}
\pagestyle{plain}

\begin{abstract}
Given the unprecedented availability of data and computing resources, there is widespread renewed interest in applying data-driven machine learning methods to problems for which the development of conventional engineering solutions is challenged by modelling or algorithmic deficiencies. This tutorial-style paper starts by addressing the questions of why and when such techniques can be useful. It then provides a high-level introduction to the basics of supervised and unsupervised learning. For both supervised and unsupervised learning, exemplifying applications to communication networks are discussed by distinguishing tasks carried out at the edge and at the cloud segments of the network at different layers of the protocol stack, with an emphasis on the physical layer. 
\end{abstract}

\section{Introduction}\label{sec:introduction}
After the ``AI winter" of the 80s and the 90s, interest in the application of data-driven Artificial Intelligence (AI) techniques has been steadily increasing in a number of engineering fields, including speech and image analysis \cite{hinton2012deep} and communications \cite{ibnkahla2000applications}. Unlike the logic-based expert systems that were dominant in the earlier work on AI (see, e.g., \cite{levesque2017common}), the renewed confidence in data-driven methods is motivated by the successes of pattern recognition tools based on machine learning. These tools rely on decades-old algorithms, such as backpropagation \cite{rumelhart1985learning}, the Expectation Maximization (EM) algorithm \cite{dempster1977maximum}, and Q-learning \cite{watkins1989learning}, with a number of modern algorithmic advances, including novel regularization techniques and adaptive learning rate schedules (see review in \cite{goodfellow2016deep}). Their success is built on the unprecedented availability of data and computing resources in many engineering domains. 

While the new wave of promises and breakthroughs around machine learning arguably falls short, at least for now, of the requirements that drove early AI research \cite{levesque2017common,pearl2018book}, learning algorithms have proven to be useful in a number of important applications -- and more is certainly on the way.

This paper provides a very brief introduction to key concepts in machine learning and to the literature on machine learning for communication systems. Unlike other review papers such as \cite{alsheikh2014machine,jiang2017machine,2018arXiv180711713Q}, the presentation aims at highlighting conditions under which the use of machine learning is justified in engineering problems, as well as specific classes of learning algorithms that are suitable for their solution. The presentation is organized around the description of general technical concepts, for which an overview of applications to communication networks is subsequently provided. These applications are chosen to exemplify general design criteria and tools and not to offer a comprehensive review of the state of the art and of the historical progression of advances on the topic.

We proceed in this section by addressing the question ``What is machine learning?", by providing a taxonomy of machine learning methods, and by finally considering the question ``When to use machine learning?''.

\begin{figure} 
	\centering
	\includegraphics[width=5.5cm]{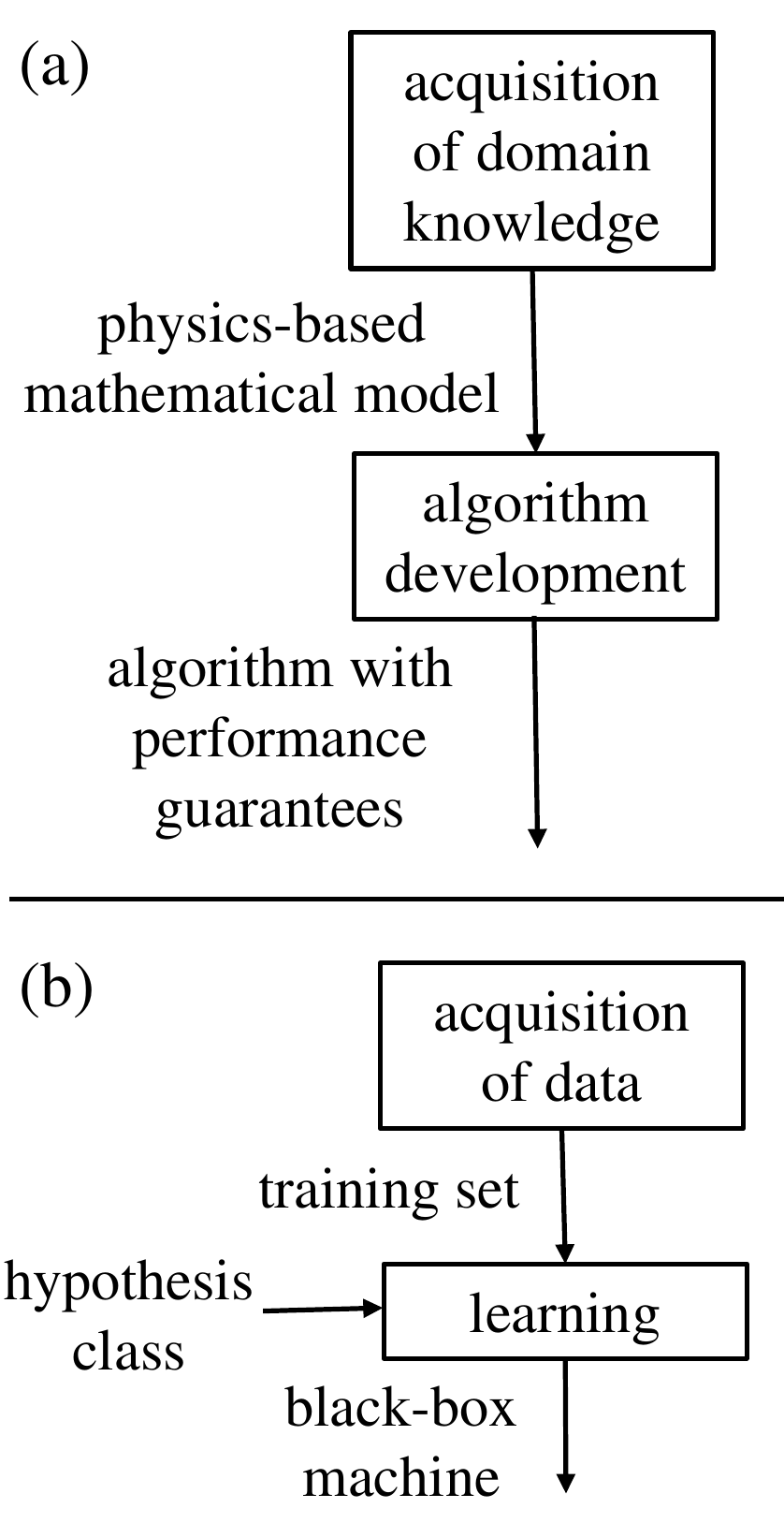}
	\caption{(a) Conventional engineering design flow; and (b) baseline machine learning methodology.}
	\label{FigMLvENG}
\end{figure}

\subsection{What is Machine Learning?}\label{sec:ia}
In order to fix the ideas, it is useful to introduce the machine learning methodology as an alternative to the conventional engineering approach for the design of an algorithmic solution. As illustrated in Fig. \ref{FigMLvENG}(a), the conventional engineering design flow starts with the \emph{acquisition of domain knowledge}: The problem of interest is studied in detail, producing a \emph{mathematical model} that capture the \emph{physics} of the set-up under study. Based on the model, an \emph{optimized algorithm} is produced that offers \emph{performance guarantees} under the assumption that the given physics-based model is an accurate representation of reality.

As an example, designing a decoding algorithm for a wireless fading channel under the conventional engineering approach would require the development, or the selection, of a physical model for the channel connecting transmitter and receiver. The solution would be obtained by tackling an optimization problem, and it would yield optimality guarantees under the given channel model. Typical example of channel models include Gaussian and fading channels (see, e.g., \cite{lin2001error}).

In contrast, in its most basic form, the machine learning approach substitutes the step of acquiring domain knowledge with the potentially easier task of collecting a sufficiently large number of examples of desired behaviour for the algorithm of interest. These examples constitute the \emph{training set}. As seen in Fig. \ref{FigMLvENG}(b), the examples in the training set are fed to a learning algorithm to produce a trained ``machine" that carries out the desired task. Learning is made possible by the choice of a set of possible ``machines", also known as the \emph{hypothesis class}, from which the learning algorithm makes a selection during training. An example of an hypothesis class is given by a neural network architecture with learnable synaptic weights. Learning algorithms are generally based on the optimization of a performance criterion that measures how well the selected ``machine" matches the available data. 


For the problem of designing a channel decoder, a machine learning approach can hence operate even in the absence of a well-established channel model. It is in fact enough to have a sufficiently large number of examples of received signals -- the inputs to the decoding machine -- and transmitted messages -- the desired outputs of the decoding machine -- to be used for the training of a given class of decoding functions \cite{gruber2017deep}. 

\begin{figure} 
	\centering
	\includegraphics[width=9cm]{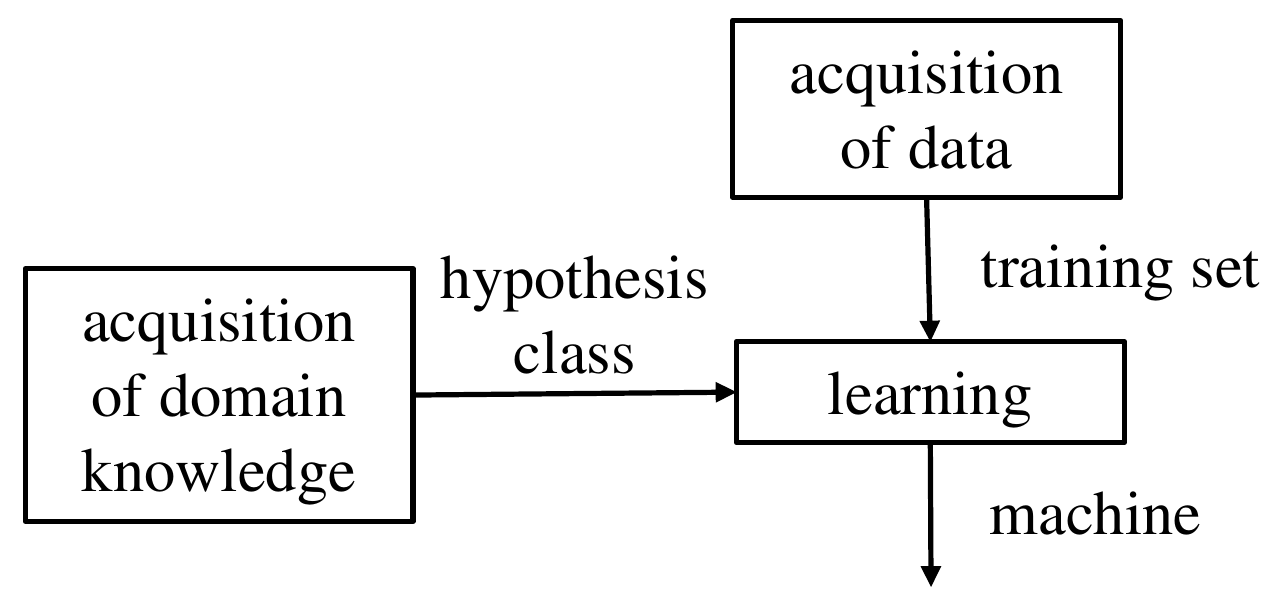}
	\caption{Machine learning methodology that integrates domain knowledge during model selection.}
	\label{FigMLvENGcon}
\end{figure}

Moving beyond the basic formulation described above, machine learning tools can \emph{integrate available domain knowledge} in the learning process. This is indeed the key to the success of machine learning tools in a number of applications. A notable example is image processing, whereby knowledge of the translational invariance of visual features is reflected in the adoption of convolutional neural networks as the hypothesis class to be trained. More generally, as illustrated in Fig. \ref{FigMLvENGcon}, domain knowledge can dictate the choice of a specific hypothesis class for use in the training process. Examples of applications of this idea to communication systems, including to the problem of decoding, will be discussed later in the paper.

\subsection{Taxonomy of Machine Learning Methods}
\begin{figure} 
	\centering
	\includegraphics[width=8cm]{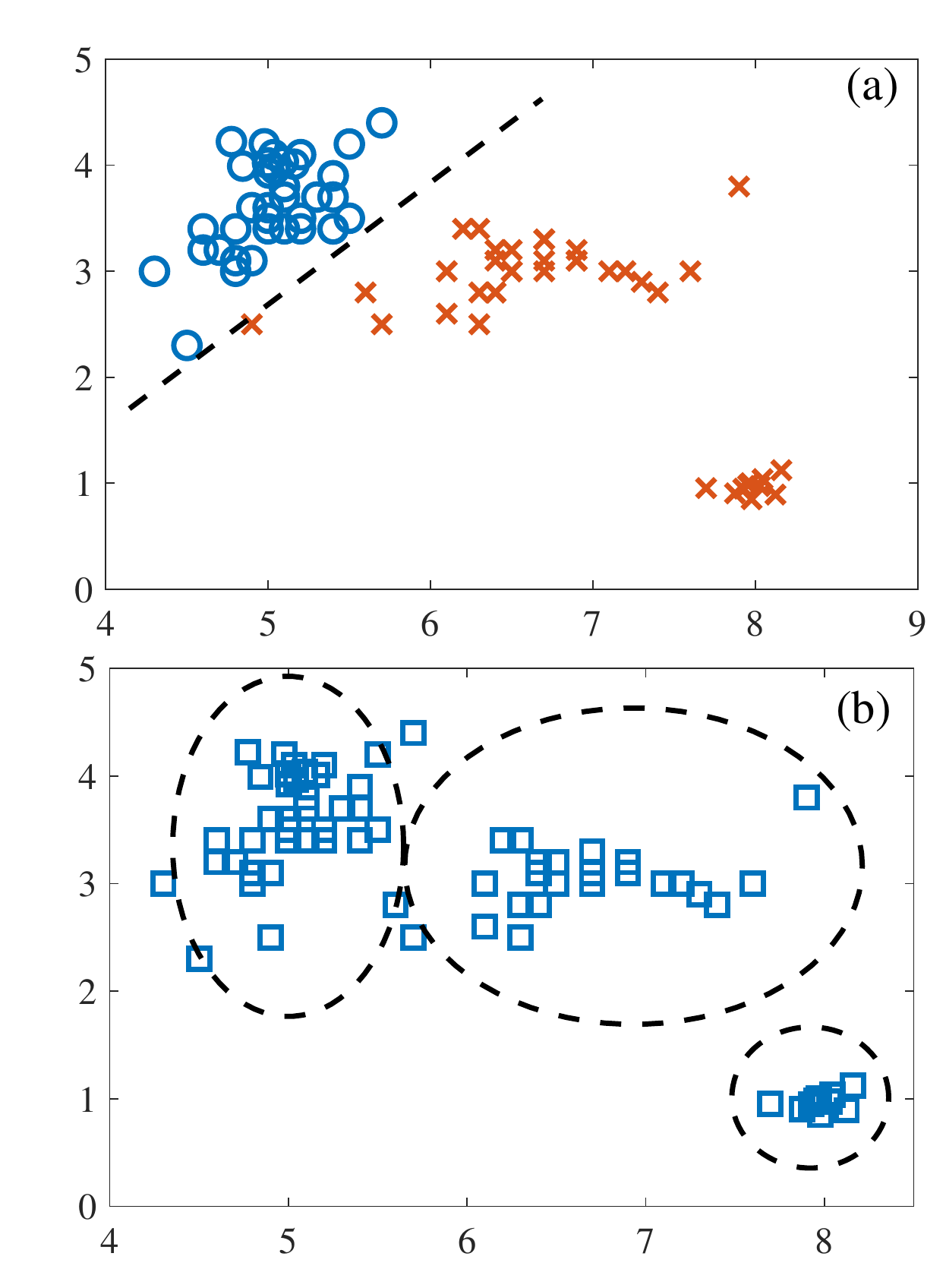}
	\caption{Illustration of (a) supervised learning and (b) unsupervised learning.}
	\label{FigXw}
\end{figure}

There are three main classes of machine learning techniques, as discussed next.
\begin{itemize}
	\item \emph{Supervised learning}: In supervised learning, the training set consists of pairs of input and desired output, and the goal is that of learning a mapping between input and output spaces. As an illustration, in Fig. \ref{FigXw}(a), the inputs are points in the two-dimensional plane, the outputs are the labels assigned to each input (circles or crosses), and the goal is to learn a binary classifier. Applications include the channel decoder discussed above, as well as email spam classification on the basis of examples of spam/ non-spam emails.  
	\item \emph{Unsupervised learning}: In unsupervised learning, the training set consists of unlabelled inputs, that is, of inputs without any assigned desired output. For instance, in Fig. \ref{FigXw}(b), the inputs are again points in the two-dimensional plane, but no indication is provided by the data about the corresponding desired output. Unsupervised learning generally aims at discovering properties of the mechanism generating the data. In the example of Fig. \ref{FigXw}(b), the goal of unsupervised learning is to cluster together input points that are close to each other, hence assigning a label -- the cluster index -- to each input point (clusters are delimited by dashed lines). Applications include clustering of documents with similar topics. It is emphasized that clustering is only one of the learning tasks that fall under the category of unsupervised learning (see Sec. \ref{sec:unsup}).
	\item \emph{Reinforcement learning}: Reinforcement learning lies, in a sense, between supervised and unsupervised learning. Unlike unsupervised learning, some form of supervision exists, but this does not come in the form of the specification of a desired output for every input in the data. Instead, a reinforcement learning algorithm receives feedback from the environment only after selecting an output for a given input or observation. The feedback indicates the degree to which the output, known as action in reinforcement learning, fulfils the goals of the learner. Reinforcement learning applies to sequential decision making problems in which the learner interacts with an environment by sequentially taking actions -- the outputs -- on the basis of its observations -- its inputs -- while receiving feedback regarding each selected action. 
\end{itemize}

Most current machine learning applications fall in the supervised learning category, and hence aim at learning an existing pattern between inputs and outputs. Supervised learning is relatively well-understood at a theoretical level \cite{shalev2014understanding,2017arXiv170605394A}, and it benefits from well-established algorithmic tools. Unsupervised learning has so far defied a unified theoretical treatment \cite{hastie2009unsupervised}. Nevertheless, it arguably poses a more fundamental practical problem in that it directly tackles the challenge of learning by direct observation without any form of explicit feedback. Reinforcement learning has found extensive applications in problems that are characterized by clear feedback signals, such as win/lose outcomes in games, and that entail searches over large trees of possible action-observation histories \cite{sutton1998reinforcement,silver2016mastering}.

This paper only covers supervised and unsupervised learning. Reinforcement learning requires a different analytical framework grounded in Markov Decision Processes and will not be discussed here (see \cite{sutton1998reinforcement}). For a broader discussion on the technical aspects of supervised and unsupervised learning, we point to \cite{simeone2017brief} and references therein.

\subsection{When to Use Machine Learning?}\label{sec:when}
Based on the discussion in Sec. \ref{sec:ia}, the use of a machine learning approach in lieu of a more conventional engineering design should be justified on a case-by-case basis on the basis of its suitability and potential advantages. The following criteria, inspired by \cite{brynjolfsson2017can}, offer useful guidelines on the type of engineering tasks that can benefit from the use of machine learning tools. 

\noindent  \textbf{1.} \emph{The traditional engineering flow is not applicable or is undesirable due to a model deficit or to an algorithm deficit} \cite{isittutorial}. 
\begin{itemize}
\item With a \emph{model deficit}, no physics-based mathematical models exist for the problem due to insufficient domain knowledge. As a result, a conventional model-based design is inapplicable. 
\item With an \emph{algorithm deficit}, a well-established mathematical model is available, but existing algorithms optimized on the basis of such model are too complex to be implemented for the given application. In this case, the use of hypothesis classes including efficient ``machines", such as neural network of limited size or with tailored hardware implementations (see, e.g., \cite{davies2018loihi,bagheri2017training} and references therein), can yield lower-complexity solutions. 
\end{itemize}
\noindent  \textbf{2.} \emph{A sufficiently large training data sets exist or can be created}.\\
\noindent  \textbf{3.} \emph{The task does not require the application of logic, common sense, or explicit reasoning based on background knowledge}.\\
\noindent  \textbf{4.} \emph{The task does not require detailed explanations for how the decision was made}. The trained machine is by and large a black box that maps inputs to outputs. As such, it does not provide direct means to ascertain why a given output has been produced in response to an input, although recent research has made some progress on this front \cite{chen2018learning}. This contrasts with engineered optimal solutions, which can be typically interpreted on the basis of physical performance criteria. For instance, a maximum likelihood decoder chooses a given output because it minimizes the probability of error under the assumed model.\\
\noindent  \textbf{5.} \emph{The phenomenon or function being learned is stationary for a sufficiently long period of time.} This is in order to enable data collection and learning.\\
\noindent  \textbf{6.} \emph{The task has either loose requirement constraints, or, in the case of an algorithm deficit, the required performance guarantees can be provided via numerical simulations}. With the conventional engineering approach, theoretical performance guarantees can be obtained that are backed by a physics-based mathematical model. These guarantees can be relied upon insofar as the model is trusted to be an accurate representation of reality. If a machine learning approach is used to address an algorithm deficit and a physics-based model is available, then numerical results may be sufficient in order to compute satisfactory performance measures. In contrast, weaker guarantees can be offered by machine learning in the absence of a physics-based model. In this case, one can provide performance bounds only under the assumptions that the hypothesis class is sufficiently general to include ``machines" that can perform well on the problem and that the data is representative of the actual data distribution to be encountered at runtime (see, e.g., \cite{simeone2017brief}[Ch. 5]). The selection of a biased hypothesis class or the use of an unrepresentative data set may hence yield strongly suboptimal performance. 

We will return to these criteria when discussing applications to communication systems.

\section{Machine Learning for Communication Networks}\label{sec:mlcomm}
In order to exemplify applications of supervised and unsupervised learning, we will offer annotated pointers to the literature on machine learning for communication systems. Rather than striving for a comprehensive, and historically minded, review, the applications and references have been selected with the goal of illustrating key aspects regarding the use of machine learning in engineering problems. 

\begin{figure} 
	\centering
	\includegraphics[width=10cm]{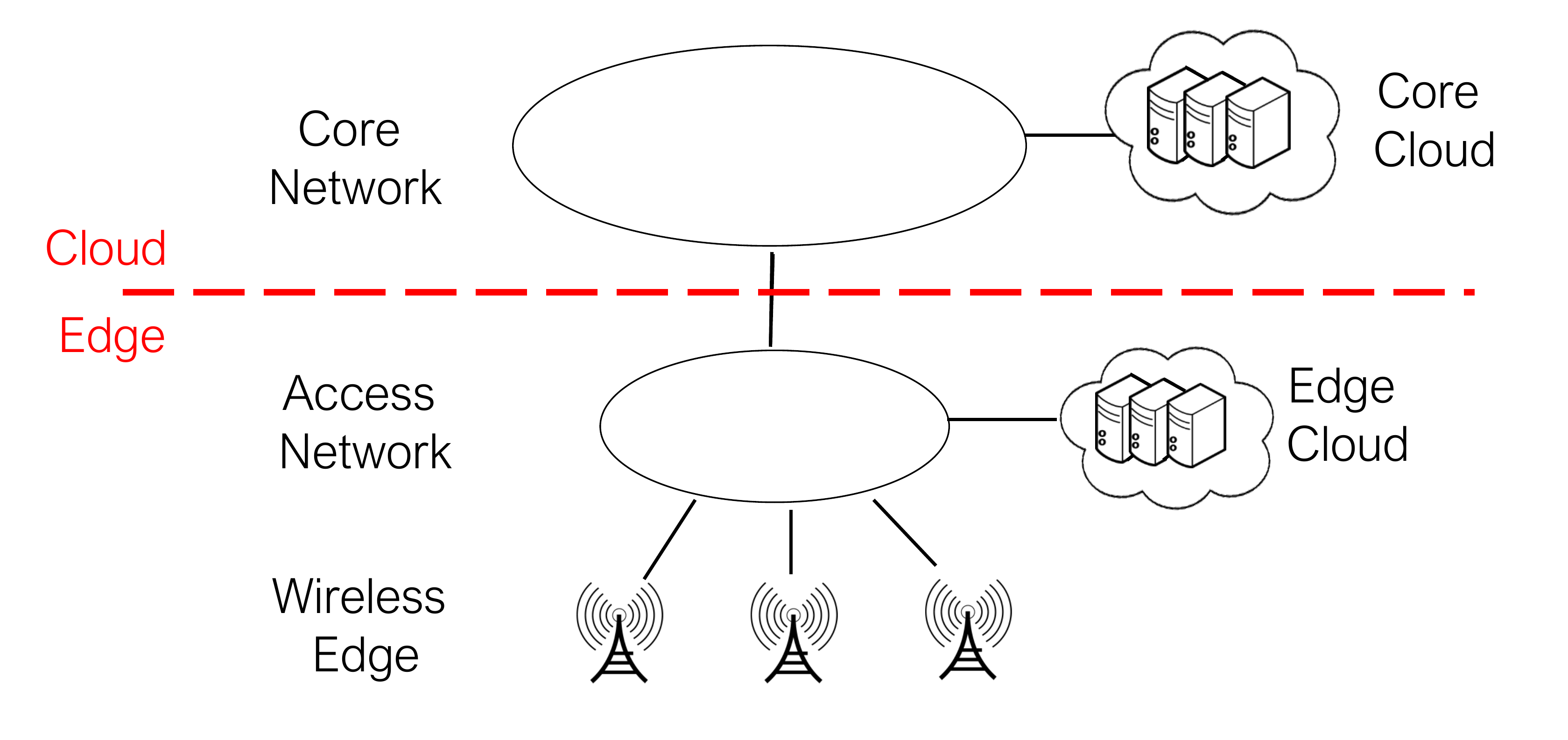}
	\caption{A generic cellular wireless network architecture that distinguishes between edge segment, with base stations, access points, and associated computing resources, and cloud segment, consisting of core network and associated cloud computing platforms.}
	\label{fogarch}
\end{figure}

Throughout, we focus on tasks carried out at the network side, rather than at the users, and organize the applications along two axes. On one, with reference to Fig. \ref{fogarch}, we distinguish tasks that are carried out at the \emph{edge} of the network, that is, at the base stations or access points and at the associated computing platforms, from tasks that are instead responsibility of a centralized \emph{cloud} processor connected to the core network (see, e.g., \cite{2018arXiv180807647P}). The edge operates on the basis of timely local information collected at different layers of the protocol stack, which may include all layers from the physical up to the application layer. In contrast, the centralized cloud processes longer-term and global information collected from multiple nodes in the edge network, which typically encompasses only the higher layers of the protocol stack, namely networking and application layers. Examples of data that may be available at the cloud and at the edge can be found in Table \ref{table:1} and Table \ref{table:2}, respectively.

\begin{table*}[t]
	\caption{Examples of data available at the edge segment of a communication network}
	\centering
	\begin{tabular}{|c | c |} 
		
		\hline
		Layer & Data\\
		\hline
		Physical & 	Baseband signals, channel state information    \\
		\hline
		Medium Access Control/ Link & Throughput, FER, random access load and
		latency     \\
		\hline
		Network & Location, traffic loads across services,
		users' device types, battery levels  \\
		\hline
		Application & Users' preferences,
		content demands, computing loads, QoS metrics   \\
		\hline
	\end{tabular}
	\label{table:1}
\end{table*}

\begin{table*}[t]
	\caption{Examples of data available at the cloud segment of a communication network}
	\centering
	\begin{tabular}{|c | c |} 
		
		\hline
		Layer & Data\\
		\hline
		Network & Mobility patterns, network-wide traffic
		statistics, outage rates  \\
		\hline
		Application &  User's behaviour patterns, subscription
		information, service usage statistics, TCP/IP traffic statistics  \\
		\hline
	\end{tabular}
	\label{table:2}
\end{table*}

As a preliminary discussion, it is useful to ask which tasks of a communication network, if any, may benefit from machine learning through the lens of the criteria reviewed in Sec. \ref{sec:when}. First, as seen, there should be either a model deficit or an algorithm deficit that prevents the use of a conventional model-based engineering design. As an example of model deficit, proactive resource allocation that is based on predictions of human behaviour, e.g., for caching popular contents, may not benefit from well-established and reliable models, making a data-driven approach desirable (see, e.g., \cite{paschos2016wireless,chen2017machine}). For an instance of algorithm deficit, consider the problem of channel decoding for channels with known and accurate models based on which the maximum likelihood decoder entails an excessive complexity.

Assuming that the problem at hand is characterized by model or algorithm deficits, one should then consider the rest of the criteria discussed in Sec. \ref{sec:when}. Most are typically satisfied by communication problems. Indeed, for most tasks in communication networks, it is possible to collect or generate training data sets and there is no need to apply common sense or to provide detailed explanations for how a decision was made. 

The remaining two criteria need to be checked on a case-by-case basis. First, the phenomenon or function being learned should not change too rapidly over time. For example, designing a channel decoder based on samples obtained from a limited number of realizations of a given propagation channel requires the channel is stationary over a sufficiently long period of time (see \cite{angjelichinoski}). 

Second, in the case of a model deficit, the task should have some tolerance for error in the sense of not requiring provable performance guarantees. For instance, the performance of a decoder trained on a channel lacking a well-established channel model, such as a biological communication link, can only be relied upon insofar as one trusts the available data to be representative of the complete set of possible realizations of the problem under study. Alternatively, under an algorithm deficit, a physics-based model, if available, can be possibly used to carry out computer simulations and obtain numerical performance guarantees. 


In Sec. \ref{sec:supappl} and Sec. \ref{sec:unsupappl}, we will provide some pointers to specific applications to supervised and unsupervised learning, respectively. 

\section{Supervised Learning}

As introduced in Sec. \ref{sec:introduction}, supervised learning aims at discovering patterns that relate inputs to outputs on the basis of a training set of input-output examples. We can distinguish two classes of supervised learning problems depending on whether the outputs are continuous or discrete variables. In the former case, we have a \emph{regression} problem, while in the latter we have a \emph{classification} problem. We discuss the respective goals of the two problems next. This is followed by a formal definition of classification and regression, and by a discussion of the methodology and of the main steps involved in tackling the two classes of problems.

\subsection{Goals}\label{sec:goals}
\begin{figure} 
	\centering
	\includegraphics[width=8.3cm]{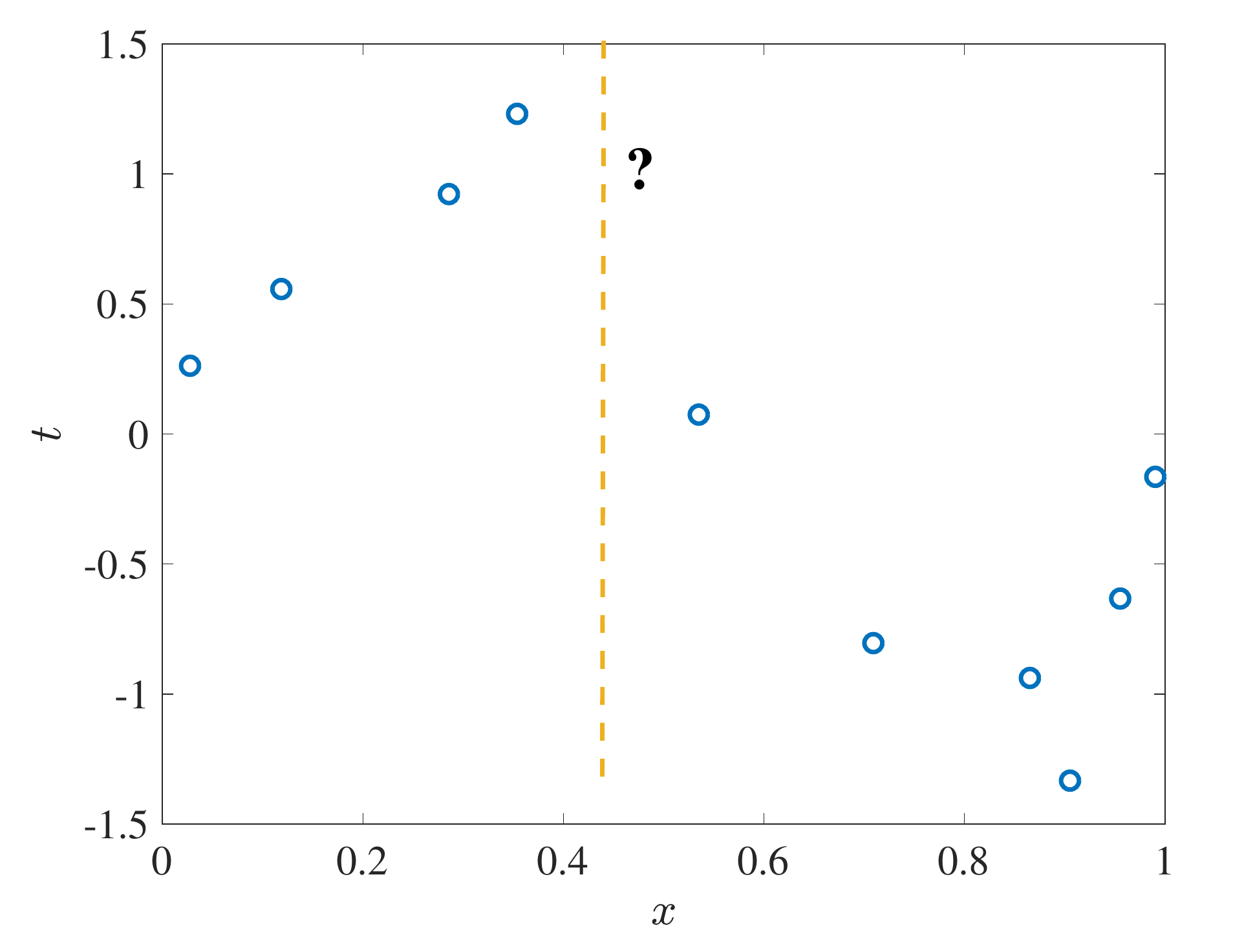}
	\caption{Illustration of the supervised learning problem of regression: Given input-output training examples $(x_{n},t_{n})$, with $n=1,...,N$, how should we predict the output $t$ for an unobserved value of the input $x$?} 
	\label{FIg1Ch2}
\end{figure}

As illustrated in Fig. \ref{FIg1Ch2}, in a regression problem, we are given a training set $\mathcal{D}$ of $N$ training points $(x_{n},t_{n})$, with
$n=1,...,N$, where the variables $x_{n}$ are the inputs, also known as covariates, domain points, or explanatory variables; while the variables $t_{n}$ are the outputs, also known as dependent variables, labels, or responses. In regression, the outputs are continuous variables.  The problem is to predict the output $t$ for a new, that is, as of yet unobserved, input $x$. 

\begin{figure} 
	\centering
	\includegraphics[width=8.3cm]{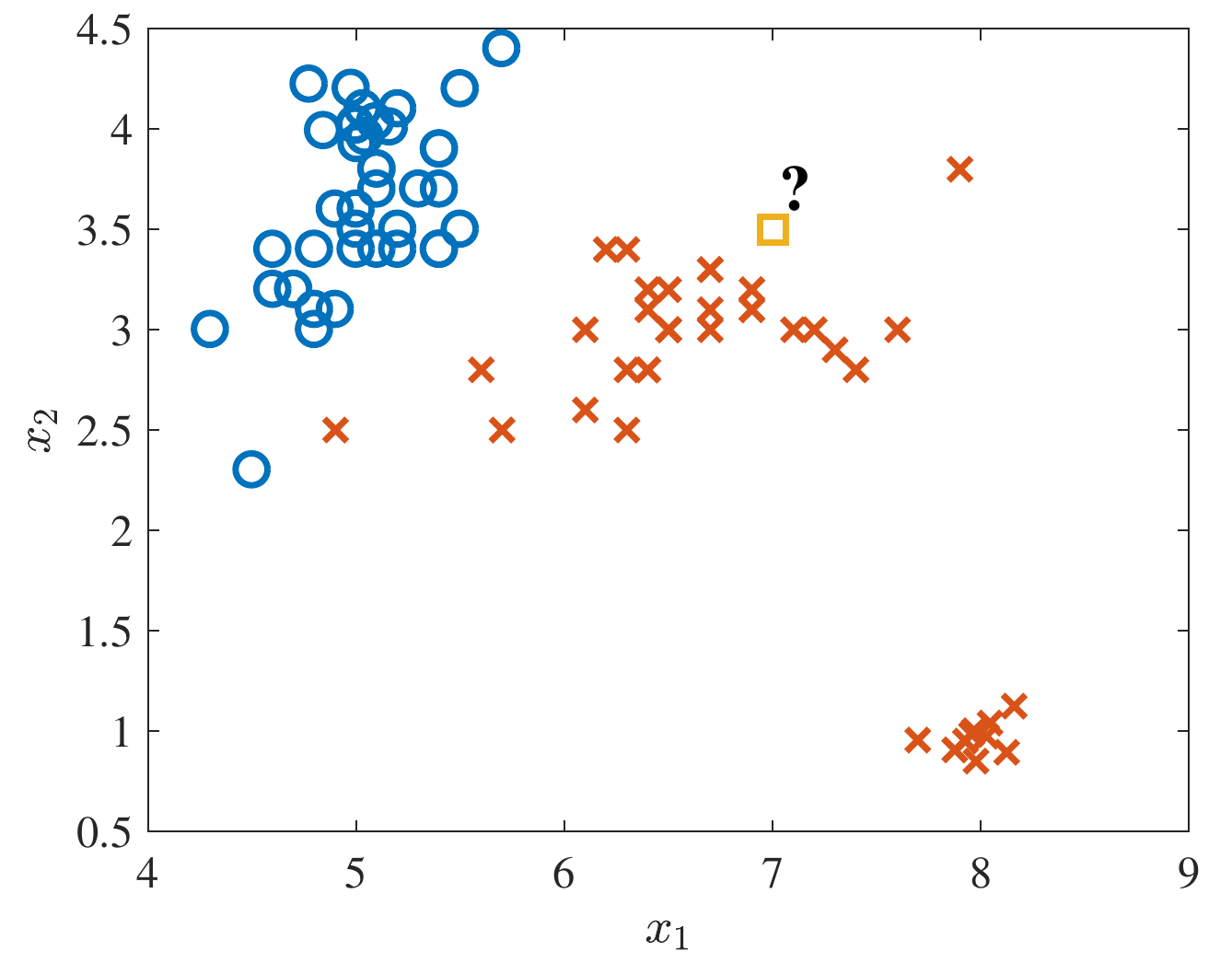}
	\caption{Illustration of the supervised learning problem of classification: Given input-output training examples $(x_{n},t_{n})$, with $n=1,...,N$, how should we predict the output $t$ for an unobserved value of the input $x$? }
		\label{Fig1Ch4}
	\end{figure}

As illustrated in Fig. \ref{Fig1Ch4}, classification is similarly defined  with the only caveat that the outputs $t$ are discrete variables that take a finite number of possible values. The value of the output $t$ for a given input $x$ indicates the class to which $x$ belongs. For instance, the label $t$ is a binary variable as in Fig. \ref{Fig1Ch4} for a binary classification problem. Based on the training set $\mathcal{D}$, the goal is to predict the label, or the class, $t$ for a new, as of yet unobserved, input $x$.

To sum up, the goal of both regression and classification is to derive from the training data set  $\mathcal{D}$ a predictor $\hat{t}(x)$ that generalizes the input-output mapping in $\mathcal{D}$ to inputs $x$ that are not present in $\mathcal{D}$. As such, learning is markedly distinct from \emph{memorizing}: while memorizing would require producing a value $t_n$ for some recorded input $x_n$ in the training set, learning is about \emph{generalization} from the data set to the rest of the relevant input space.

The problem of extrapolating a predictor from the training set is evidently impossible unless one is willing to make some assumption about the underlying input-output mapping. In fact, the output $t$ may well equal any value for an unobserved $x$ if nothing else is specified about the problem. This impossibility is formalized by the \emph{no free-lunch theorem}: without making assumptions about the relationship between input and output, it is not possible to generalize the available observations outside the training set \cite{shalev2014understanding}. The set of assumptions made in order to enable learning are known as \emph{inductive bias}. As an example, for the regression problem in Fig. \ref{FIg1Ch2}, a possible inductive bias is to postulate that the input-output mapping is a polynomial function of some order. 

\subsection{Defining Supervised Learning}\label{sec:defsup}

Having introduced the goal of supervised learning, we now provide a more formal definition of the problem. Throughout, we use Roman font to denote random variables and the corresponding letter in regular font for realizations.

As a starting point, we assume that the training set $\mathcal{D}$ is generated as  
\begin{equation}
(\mathrm{x}_{n},\mathrm{t}_{n})\underset{\textrm{i.i.d.}}{\sim}p(x,t),\textrm{ \ensuremath{n=1,...,N}},
\end{equation}
that is, each training sample pair $(\mathrm{x}_{n},\mathrm{t}_{n})$ is generated from the same true joint distribution $p(x,t)$ and the sample pairs are independent identically distributed (i.i.d.). As discussed, based on the training set $\mathcal{D}$, we wish to obtain a predictor $\hat{t}(x)$ that performs well on any possible relevant input $x$. This requirement is formalized by imposing that the predictor is accurate for any \emph{test pair} $(\mathrm{x},\mathrm{t})\sim p(x,t)$, which is generated independently of all the pairs in the training set $\mathcal{D}$. 

The quality of the prediction $\hat{t}(x)$ for a test pair ($x,t$) is measured by a given loss function $\ell(t,\hat{t})$ as $\ell(t,\hat{t}(x))$. Typical examples of loss functions include the quadratic loss $\ell(t,\hat{t})=(t-\hat{t})^{2}$ for regression problems; and the error rate $\ell(t,\hat{t})=1(t\neq\hat{t})$, which equals 1 when the prediction is incorrect, i.e., $t\neq\hat{t}$, and 0 otherwise, for classification problems. 

The formal goal of learning is that of minimizing the average loss on the test pair, which is referred to as the \emph{generalization loss}. For a given predictor $\hat{t}$, this is defined as
\begin{equation}\label{eq:genloss}
L_{p}(\hat{t})=\textrm{E}_{(\mathrm{x},\mathrm{t})\sim p(x,t)}[\ell(\mathrm{t},\hat{t}(\mathrm{x}))].
\end{equation}
The generalization loss (\ref{eq:genloss}) is averaged over the distribution of the test pair ($x,t$). 

Before moving on to the solution of the problem of minimizing the generalization loss, we mention that the formulation provided here is only one, albeit arguably the most popular, of a number of alternative formulations of supervised learning. The frequentist framework described above is in fact complemented by other viewpoints, including Bayesian and Minimum Description Length (MDL) (see \cite{simeone2017brief} and references therein).

\subsection{When The True Distribution $p(x,t)$ is Known: Inference}\label{sec:inference}
	
Consider first the case in which the true joint distribution $p(x,t)$ relating input and output is known. This scenario can be considered as an idealization of the situation resulting from the conventional engineering design flow when the available physics-based model is accurate (see Sec. \ref{sec:introduction}). Under this assumption, the data set $\mathcal{D}$ is not necessary, since the mapping between input and output is fully described by the distribution $p(x,t)$. 

If the true distribution $p(x,t)$ is known, the problem of minimizing the generalization loss reduces to a standard \emph{inference} problem, i.e., an estimation problem in a regression set-up, in which the outputs are continuous variables, or a detection problem in a classification set-up, in which the outputs are finite discrete variables. 

In an inference problem, the optimal predictor $\hat{t}$ can be directly computed from the \emph{posterior} distribution \begin{equation}\label{eq:posterior} p(t|x)=\frac{p(x,t)}{p(x)}, \end{equation} where $p(x)$ is the marginal distribution of the input $\mathrm{x}$. The latter can be computed from the joint distribution $p(x,t)$ by summing or integrating out all the values of $t$. In fact, given a loss function $\ell(t,\hat{t})$, the optimal predictor for any input $x$ is obtained as \begin{equation}\label{eq:optpred}
	\hat{t}^{*}(x)=\textrm{arg}\min_{\hat{t}}\textrm{E}_{\mathrm{t}\sim p(t|x)}[\ell(\mathrm{t},\hat{t})|x].
\end{equation} In words, the optimal predictor $\hat{t}^{*}(x)$ is obtained by identifying the value (or values) of $t$ that minimizes the average loss, where the average is taken with respect to the posterior distribution $p(t|x)$ of the output given the input. Given that the posterior $p(t|x)$ yields the optimal predictor, it is also known as the \emph{true predictive distribution}.

The optimal predictor (\ref{eq:optpred}) can be explicitly evaluated for given loss functions. For instance, for the quadratic loss, which is typical for regression, the optimal predictor is given by the mean of the predictive distribution, or the posterior mean, i.e., \begin{equation}
\hat{t}^{*}(x)=\textrm{E}_{\mathrm{t}\sim p(t|x)}[\mathrm{t}|x],\end{equation} while, with the error rate loss, which is typical for classification, problems, the optimal predictor is given by the maximum of the predictive distribution, or the maximum a posteriori (MAP) estimate, i.e.,\begin{equation}
 \hat{t}^{*}(x)=\textrm{arg}\max_{t}p(t|x). \end{equation}

For a numerical example, consider binary inputs and outputs and the joint distribution $p(x,t)$ such that $p(0,0)=0.05$, $p(0,1)=0.45$, $p(1,0)=0.4$ and $p(1,1)=0.1$. The predictive distribution for input $x=0$ is then given as $p(t=1|x=0)=0.9$, and hence we have the optimal predictor given by the average $\hat{t}^{*}(x=0)=0.9\times1+0.1\times0=0.9$ for the quadratic loss, and by the MAP solution $\hat{t}^{*}(x=0)=1$ for the error rate loss.

\subsection{When the True Distribution $p(x,t)$ is Not Known: Machine Learning}\label{sec:notknown}

Consider now the case of interest in which domain knowledge is not available and hence the true joint distribution is unknown. In such a scenario, we have a learning problem and we need to use the examples in the training set $\mathcal{D}$ in order to obtain a meaningful predictor that approximately minimizes the generalization loss. At a high level, the methodology applied by machine learning follows three main steps, which are described next.

1. \emph{Model selection (inductive bias)}: As a first step, one needs to commit to a specific class of \emph{hypotheses} that the learning algorithm may choose from. The hypothesis class is also referred to as model. The selection of the hypothesis class characterizes the inductive bias mentioned above as a pre-requisite for learning. In a probabilistic framework, the hypothesis class, or model, is defined by a family of probability distributions parameterized by a vector $\theta$. Specifically, there are two main ways of specifying a parametric family of distributions as a model for supervised learning: \begin{itemize}
	\item \emph{Generative model}: Generative models specify a family of joint distributions $p(x,t|\theta)$;
	\item \emph{Discriminative model}: Discriminative models parameterize directly the predictive distribution as $p(t|x,\theta)$. 
	\end{itemize} Broadly speaking, discriminative models do not make any assumptions about the distribution of the inputs $x$ and hence may be less prone to bias caused by a misspecification of the hypothesis class. On the flip side, generative models may be able to capture more of the structure present in the data and consequently improve the performance of the predictor \cite{seeger2006taxonomy}. For both types of models, the hypothesis class is typically selected from a common set of probability distributions that lead to efficient learning algorithms in Step 2. Furthermore, any available basic domain knowledge can be in principle incorporated in the selection of the model (see also Sec. \ref{sec:conclusions}).

2. \emph{Learning}: Given data $\mathcal{D}$, in the learning step, a learning criterion is optimized in order to obtain the parameter vector $\theta$ and identify a distribution $p(x,t|\theta)$ or $p(t|x,\theta)$, depending on whether a generative or discriminative model was selected at Step 1. 

3. \emph{Inference:} In the inference step, the learned model is used to obtain the predictor $\hat{t}(x)$ by using (\ref{eq:optpred}) with the learned model in lieu of the true distribution. Note that generative models require the calculation of the predictive distribution $p(t|x)$ via marginalization, while discriminative models provide directly the predictive distribution. As mentioned, the predictor should be evaluated on test data that is different from the training set $\mathcal{D}$. As we will discuss, the design cycle typically entails a loop between validation of the predictor at Step 3 and model selection at Step 1.

The next examples illustrate the three steps introduced above for a binary classification problem. 

\begin{figure} 
	\centering
	\includegraphics[width=8.3cm]{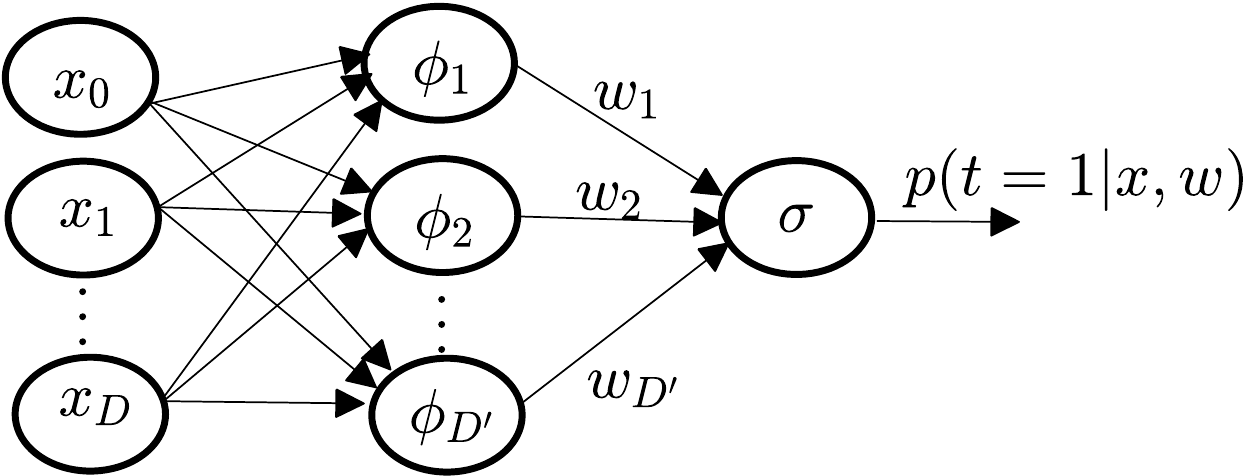}
	\caption{An illustration of the hypothesis class $p(t|x,w)$ assumed by logistic regression using a neural network representation: functions $\phi_i$, with $i=1,...,D'$, are fixed and compute features of the input vector $x=[x_1,...,x_D]$. The learnable parameter vector $\theta$ here corresponds to the weights $w$ used to linearly combine the features in (\ref{eq:logregr}).}
	\label{FigNNCh42}
\end{figure}

\emph{Example 1}: Consider a binary classification problem in which the input is a generic $D$-dimensional vector $x=[x_1,...,x_D]^T$ and the output is binary, i.e., $t\in\{0,1\}$. The superscript ``$T$" represents transposition. In Step 1, we select a model, that is, a parameterized family of distributions. A common choice is given by logistic regression\footnote{The term "regression" may be confusing, since the model applies to classification.}, which is a discriminative model whereby the predictive distribution $p(t|x,\theta)$ is parameterized as illustrated in Fig. \ref{FigNNCh42}. The model first computes $D'$ fixed features $\phi(x)=[\phi_{1}(x)\cdots\phi_{D'}(x)]^{T}$ of the input, where a feature is a function of the data. Then, it computes the predictive probability as \begin{equation}\label{eq:logregr}
p(t=1\vert x,w)=\sigma(w^{T}\phi(x)),
\end{equation}
where $w$ is the set of learnable weights -- i.e., the parameter $\theta$ defined above -- and $\sigma(a)=(1+\exp(-a))^{-1}$ is the sigmoid function. 

Under logistic regression, the probability that the label is $t=1$ increases as the linear combination of features becomes more positive, and we have $p(t=1\vert x,w)>0.5$ for $w^{T}\phi(x)>0$. Conversely, the probability that the label is $t=0$ increases as the linear combination of features becomes more negative, with $p(t=0\vert x,w)>0.5$ for $w^{T}\phi(x)<0$. As a specific instance of this problem, if we wish to classify emails between spam and non-spam ones, possible useful features may count the number of times that certain suspicious words appear in the text.

Step 2 amounts to the identification of the weight vector $w$ on the basis of the training set $\mathcal{D}$ with the ideal goal of minimizing the generalization loss (\ref{eq:genloss}). This step will be further discussed in the next subsection. Finally, in Step 3, the optimal predictor is obtained by assuming that the learned model $p(t\vert x,w)$ is the true predictive distribution. Assuming an error rate loss function, following the discussion in Sec. \ref{sec:inference}, the optimal predictor is given by the MAP choice $\hat{t}^{*}(x)=1$ if $w^{T}\phi(x)>0$ and $\hat{t}^{*}(x)=0$ otherwise. It is noted that the linear combination $w^{T}\phi(x)$ is also known as \emph{logit or log-likelihood ratio (LLR)}. This rule can be seen to correspond to a \emph{linear classifier} \cite{simeone2017brief}. The performance of the predictor should be tested on new, test, input-output pairs, e.g., new emails in the spam classification example. $\square$

\begin{figure} 
	\centering
	\includegraphics[width=8.3cm]{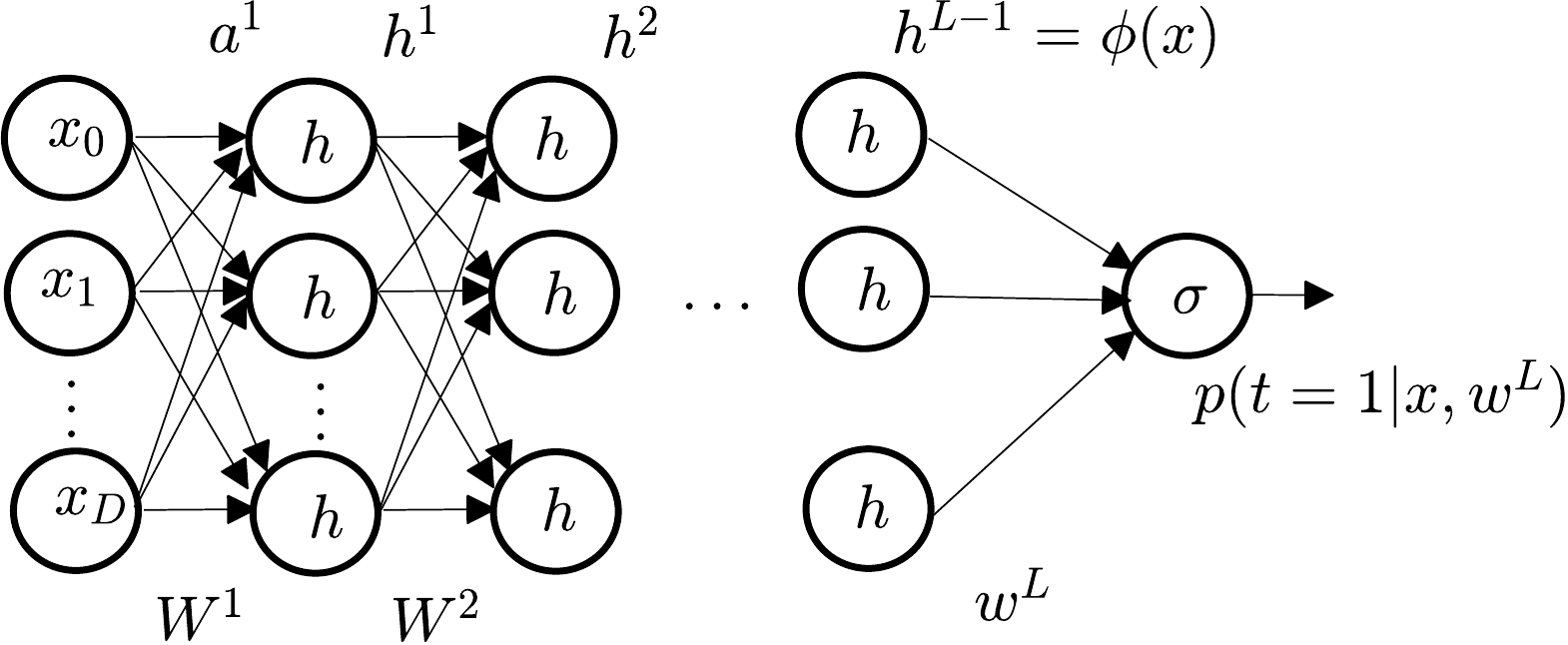}
	\caption{An illustration of the hypothesis class $p(t|x,w)$ assumed by multi-layer neural networks. The learnable parameter vector $\theta$ here corresponds to the weights $w^L$ used at the last layer to linearly combine the features $\phi(x)$ and the weight matrices $W^1,...,W^{L-1}$ used at the preceding layers in order to compute the feature vector.}
	\label{FigMultiLsigm}
\end{figure}

\emph{Example 2}: Logistic regression requires to specify a suitable vector of features $\phi(x)$. As seen in the email spam classification example, this entails the availability of some domain knowledge to be able to ascertain which functions of the input $x$ may be more relevant for the classification task at hand. As discussed in Sec. \ref{sec:introduction}, this knowledge may not be available due to, e.g., cost or time constraints. \emph{Multi-layer neural networks} provide an alternative model choice at Step 1 that obviates the need for hand-crafted features. The model is illustrated in Fig. \ref{FigMultiLsigm}. Unlike linear regression, in a multi-layer neural network, the feature vector $\phi(x)$ used by the last layer to compute the logit, or LLR, that determines the predictive probability (\ref{eq:logregr}) is not fixed a priori. Rather, the feature vector is computed by the previous layers. To this end, each neuron, represented as a circle in Fig. \ref{FigMultiLsigm}, computes a fixed non-linear function, e.g., sigmoid, of a linear combination of the values obtained from the previous layer. The weights of these linear combinations are part of the learnable parameters $\theta$, along with the weights of the last layer. By allowing the weights at all layers of the model to be trained simultaneously, multi-layer neural networks enable the joint learning of the last-layer linear classifier and of the features $\phi(x)$ the classifier operates on.  As a notable example, deep neural networks are characterized by a large number of intermediate layers that tend to learn increasingly abstract features of the input \cite{goodfellow2016deep}. $\square$

In the rest of this section, we first provide some technical details about Step 2, i.e., learning, and then we return to Step 1, i.e., model selection. As it will be seen, this order is dictated by the fact that model selection requires some understanding of the learning process.

\subsection{Learning}\label{sec:learning}

Ideally, a learning rule should obtain a predictor that minimizes the generalization error (\ref{eq:genloss}). However, as discussed in Sec. \ref{sec:inference}, this task is out of reach without knowledge of the true joint distribution $p(x,t)$. Therefore, alternative learning criteria need to be considered that rely on the training set $\mathcal{D}$ rather than on the true distribution. 

In the context of probabilistic models, the most basic learning criterion is Maximum Likelihood (ML). ML selects a value of $\theta$ in the parameterized family of models $p(x,t|\theta)$ or $p(t|x,\theta)$ that is the most likely to have generated the observed training set $\mathcal{D}$. Mathematically, ML solves the problem of maximizing the log-likelihood function \begin{equation}
\label{eq:MLprob}
	\text{maximize }\ln p(\mathcal{D}|\theta)
\end{equation} over $\theta$, where $p(\mathcal{D}|\theta)$ is the probability of the data set $\mathcal{D}$ for a given value of $\theta$. Given the assumption of i.i.d. data points in $\mathcal{D}$ (see Sec. \ref{sec:defsup}), the log-likelihood can be written as
	\begin{equation}
\ln p(\mathcal{D}|\theta)=\sum_{n=1}^{N}\ln p(t_{n}|x_{n},\theta),
	\end{equation}
where we have used as an example the case of discriminative models. Note that most learning criteria used in practice can be interpreted as ML problems, including the least squares criterion -- ML for Gaussian models -- and cross-entropy -- ML for categorical models.

The ML problem (\ref{eq:MLprob}) rarely has analytical solutions and is typically addressed by Stochastic Gradient Descent (SGD). Accordingly, at each iteration, subsets of examples, also known as \emph{mini-batches}, are selected from the training set, and the parameter vector is updated in the direction of gradient of the log-likelihood function as evaluated on these examples. The resulting learning rule can be written as\begin{equation}\label{eq:SGD}
\theta^{\textrm{new}}\leftarrow\theta^{\textrm{old}}+\gamma\nabla_{\theta}\ln p(t_{n}|x_{n},\theta)|_{\theta=\theta^{\textrm{old}}},
\end{equation} where  we have defined as $\gamma>0$ the learning rate, and, for simplicity of notation, we have considered a mini-batch given by a single example $(x_n,t_n)$. It is noted that, with multi-layer neural networks, the computation of the gradient $\nabla_{\theta}\ln p(t_{n}|x_{n},\theta)$ yields the standard backpropagation algorithm \cite{goodfellow2016deep,simeone2017brief}. The learning rate is an example of \emph{hyperparameters} that define the learning algorithm. Many variations of SGD have been proposed that aim at improving convergence (see, e.g., \cite{goodfellow2016deep,simeone2017brief}).

ML has evident drawbacks as an indirect means of minimizing the generalization error. In fact, ML only considers the fit of the probabilistic model on the training set without any consideration for the performance on unobserved input-output pairs. This weakness can be somewhat mitigated by \emph{regularization} \cite{goodfellow2016deep,simeone2017brief} during learning and by a proper selection of the model via validation, as discussed in the next subsection. Regularization adds a penalty term to the log-likelihood that depends on the model parameters $\theta$. The goal is to prevent the learned model parameters $\theta$ to assume values that are a priori too unlikely and that are hence possible symptoms of overfitting. As an example, for logistic regression, one can add a penalty that is proportional to the norm $||w||^2$ of the weight vector $w$ in order to prevent the weights to assume excessively high values when fitting the data in the learning step. 

\subsection{Model Selection}
We now discuss the first, key, step of model selection, which defines the inductive bias adopted in the learning process. In order to illustrate the main ideas, here we study a particular aspect of model selection, namely that of \emph{model order selection}. To this end, we consider a hierarchical set of models of increasing complexity and we address the problem of selecting (in Step 1) the order, or the complexity, of the specific model to be posited for learning (in Step 2). As an example of model order selection, one may fix a set of models including multi-layer networks of varying number of intermediate layers and focus on determining the number of layers. It is emphasized that the scope of model selection goes much beyond model order selection, including the possible incorporation of domain knowledge and the tuning of the hyperparameters of the learning algorithm.

\begin{figure} 
	\centering
	\includegraphics[width=8.3cm]{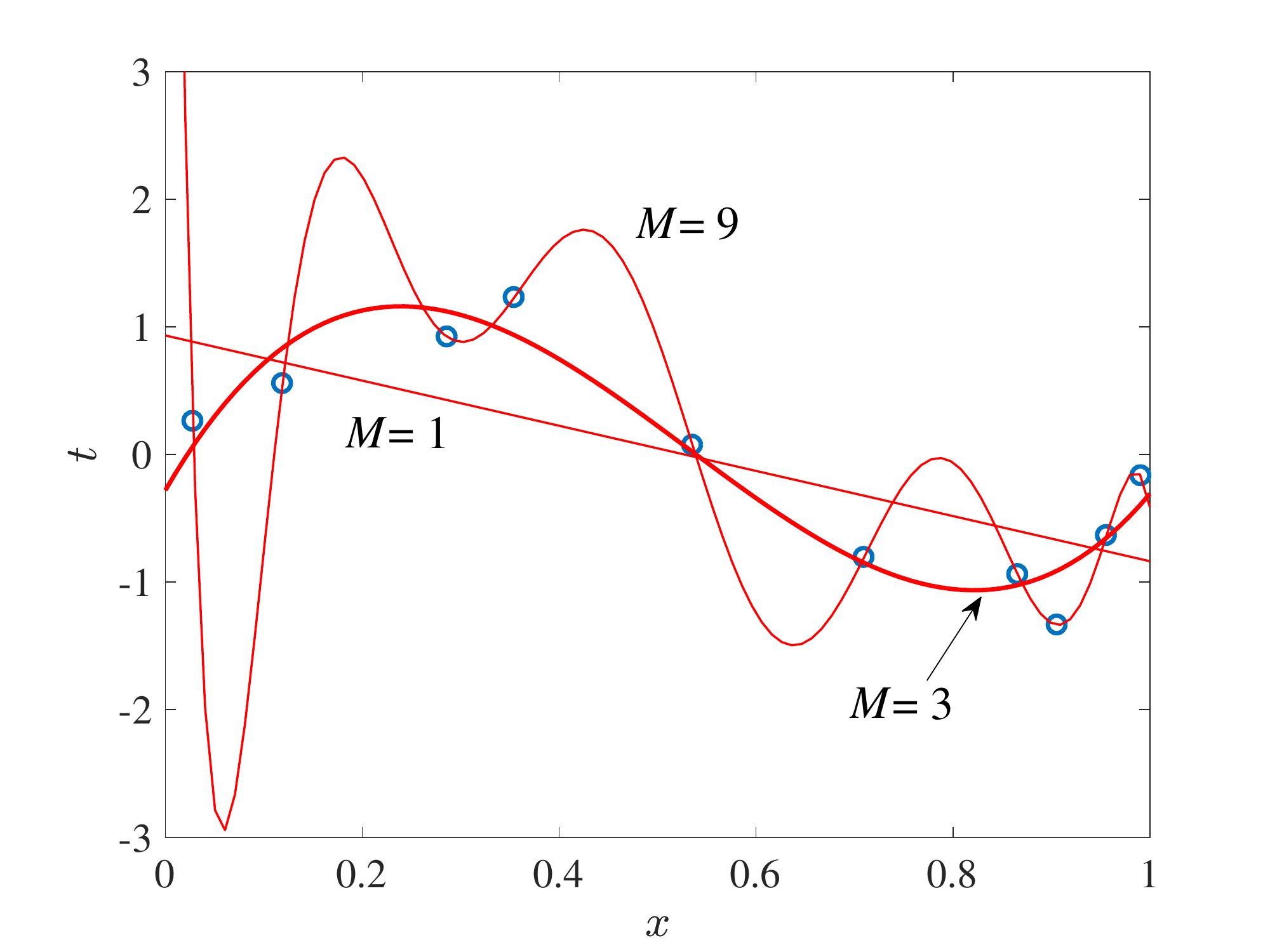}
	\caption{Training set in Fig. \ref{FIg1Ch2}, along with a predictor trained by using the discriminative model (\ref{eq:polymodel}) and ML for different values of the model order $M$.}
	\label{FIg2Ch23}
\end{figure}

For concreteness, we focus on the regression problem illustrated in Fig. \ref{FIg1Ch2} and assume a set of discriminative models $p(t|x,w)$ under which the output $t$ is distributed as\begin{equation}\label{eq:polymodel} \sum_{m=0}^{M}w_{m}x^{m}+\mathcal{N}(0,1).
\end{equation} In words, the output $t$ is given by a polynomial function of order $M$ of the input $x$ plus zero-mean Gaussian noise of power equal to one. The learnable parameter vector $\theta$ is given by the weights $w=[w_0,...,w_{M-1}]^T$. Model selection, to be carried out in Step 1, amounts to the choice of the model order $M$. 

Having chosen $M$ in Step 1, the weights $w$ can be learned in Step 2 using ML, and then the optimal predictor can be obtained for inference in Step 3. Assuming the quadratic loss, the optimal predictor is given by the posterior mean $\hat{t}(x)=\sum_{m=0}^{M}w_{m}x^{m}$ for the learned parameters $w$. This predictor is plotted in Fig. \ref{FIg2Ch23} for different values of $M$, along with the training set of Fig. \ref{FIg1Ch2}.

With $M=1$, the predictor $\hat{t}(x)$ is seen to \emph{underfit} the training data. This is in the sense that the model is not rich enough to capture the variations present in the training data, and, as a result, we obtain a large \emph{training loss}\begin{equation}\label{eq:trloss}L_{\mathcal{D}}(w)=\frac{1}{N}\sum_{n=1}^{N}(t_{n}-\hat{t}(x_{n}))^{2}.
\end{equation} The training loss measures the quality of the predictor defined by weights $w$ on the points in the training set. In contrast, with $M=9$, the predictor fits well the training data -- so much so that it appears to \emph{overfit} it. In other words, the model is too rich and, in order to account for the observations in the training set, it \emph{appears} to yield inaccurate predictions outside it. As a compromise between underfitting and overfitting, the selection $M=3$ seems to be preferable.

As implied by the discussion above, underfitting can be detected by observing solely the training data $\mathcal{D}$ via the evaluation of the training loss (\ref{eq:trloss}). In contrast, overfitting cannot be ascertained on the basis of the training data as it refers to the performance of the predictor outside $\mathcal{D}$. It follows that model selection cannot be carried out by observing only the training set. Rather, some information must be available about the generalization performance of the predictor. This is typically obtained by means of \emph{validation}. In its simplest instantiation, validation partitions the available data into two sets, a training set $\mathcal{D}$ and a \emph{validation set}. The training set is used for learning as discussed in Sec. \ref{sec:learning}, while the validation set is used to estimate the generalization loss. This is done by computing the average in (\ref{eq:trloss}) only over the validation set. More sophisticated forms of validation exist, including cross-validation \cite{goodfellow2016deep}. 

\begin{figure} 
	\centering
	\includegraphics[width=8.3cm]{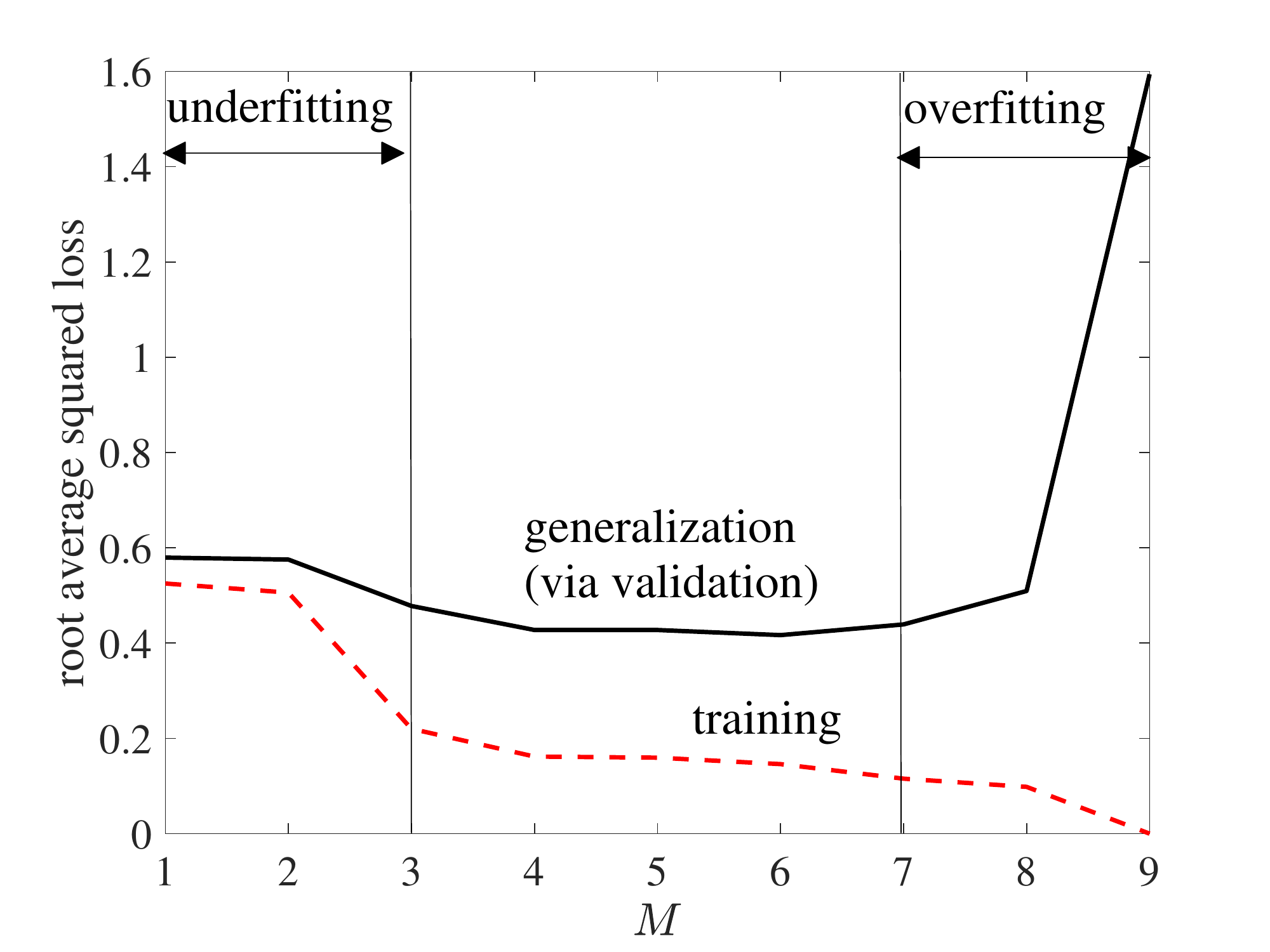}
	\caption{Training loss and generalization loss, estimated via validation, as a function of the model order $M$ for the example in Fig. \ref{FIg2Ch23}.}
	\label{Fig3Ch2}
\end{figure}

Keeping some data aside for validation, one can obtain a plot as in Fig. \ref{Fig3Ch2}, where the training loss (\ref{eq:trloss}) is compared with the generalization loss (\ref{eq:genloss}) estimated via validation. The figure allows us to conclude that, when $M$ is large enough, the generalization loss starts increasing, indicating overfitting. Note, in contrast, that underfitting is detectable by observing the training loss. A figure such as Fig. \ref{Fig3Ch2} can be used to choose a value of $M$ that approximately minimizes the generalization loss. 

More generally, validation allows for model selection, as well as for the selection of the parameters used by learning the algorithm, such as the learning rate $\gamma$ in (\ref{eq:SGD}). To this end, one compares the generalization loss, estimated via validation, for a number of models and then chooses the one with the smallest estimated generalization loss. 

Finally, it is important to remark that the performance of the model selected via validation should be estimated on the basis of a separate data set, typically called the \emph{test set}. This is because the generalization loss estimated using validation is a biased estimate of the true generalization loss (\ref{eq:genloss}) due to the process of model selection. In particular, the loss on the validation set will tend to be small, since the model was selected during validation with the aim of minimizing it. Importantly, the test set should never be used during the three steps that make up the machine learning methodology and should ideally only be used once to test the trained predictor.

\section{Applications of Supervised Learning to Communication Systems}\label{sec:supappl}
In this section, we provide some pointers to existing applications of supervised learning to communication networks. The discussion is organized by following the approach described in Sec. \ref{sec:mlcomm}. Accordingly, we distinguish between tasks carried out at edge and cloud (see Fig. \ref{fogarch}), as well as at different layers of the protocol stack. We refer to Table \ref{table:1} and Table \ref{table:2} for examples of data types that may be available at the edge and cloud segments.

\subsection{At the Edge}
Consider first tasks to be carried out at the edge, i.e., at the base stations or at the associated edge computing platform. 

\subsubsection{Physical Layer}
For the physical layer, we focus first on the receiver side and then on the transmitter. At the \emph{receiver}, a central task that can potentially benefit from machine learning is \emph{channel detection and decoding}. This amounts to a multi-class classification problem, in which the input $x$ is given by the received baseband signal and the output is the label of the correct transmitted message (e.g., the transmitted bits) \cite{gruber2017deep,o2017introduction}. When can machine learning help? Recalling the discussion in Sec. \ref{sec:mlcomm}, we should first ask whether a modelling or algorithmic deficit exists. A model deficit may occur when operating over channels that do not have well-established mathematical models, such as for molecular communications \cite{farsad2018neural}. Algorithm deficit is more common, given that optimal decoders over a number of well-established channel models tend to be computationally complex. This is the case for channels with strong non-linearities, as recognized as early as the nineties in the context of satellite communication \cite{bouchired1998equalisation,ibnkahla2000applications} and more recently for optical communications \cite{wang2006supervised}; or for modulation schemes such as continuous phase modulation \cite{de1992neural} -- another work from the nineties -- or in multi–user networks \cite{2018arXiv180709571J}. 

Assuming that the problem at hand is characterized by a modelling or algorithmic deficit, then one should also check the remaining criteria listed in Sec. \ref{sec:mlcomm}, particularly those regarding the rate of change of the phenomenon under study and the requirements in terms of performance guarantees. For channel decoding, the presence of fast-varying channels may make the first criterion hard to be satisfied in practice (unless channel estimation is made part of the learning process); while stringent reliability requirements may preclude the use of machine learning in the presence of a model deficit. 

As mentioned, a generally beneficial idea in the use of data-aided methods is that of \emph{incorporating domain knowledge in the definition of the hypothesis class}. As notable examples related to channel decoding, in \cite{nachmani2016learning,lugosch2017neural}, knowledge of the near-optimality of message passing methods for the decoding of sparse graphical codes is used to set up a parameterized model that borrows the message passing structure and that is trained to decode more general codes. A related approach is investigated in \cite{cammerer2017scaling} for polar codes. 

Another useful idea is that of directly integrating algorithms designed using the standard engineering flow with trained machines. Instances of this idea include \cite{schibisch2018online} in which a conventional channel decoder is deployed in tandem with a channel equalizer at its input that is trained to compensate for hardware impairments. A related approach is proposed in  \cite{bpcnn}, whereby a conventional decoder is implemented within a turbo-like iterative loop with a machine learning-based regressor that has the role of estimating the channel noise.

Other tasks that can potentially benefit from machine learning at the receiver's side include modulation classification, which is a classification problem justified by the complexity of optimal solutions (algorithm deficit) \cite{agirman2011modulation}; localization, which is a regression problem, typically motivated by the lack of tractable channels for complex propagation environments (model deficit) \cite{fang2008indoor}; and channel state information-based authentication, a classification problem made difficult by the absence of well-established models relating channel features with devices' identities (model deficit) \cite{2018arXiv180709469W}.

Turning to the \emph{transmitter} side, most emerging applications tackle the algorithmic deficit related to the complexity of the non-convex programs that typically underlie power control and precoding optimization for the downlink. Notably, in \cite{sun2017learning}, a training set is obtained by running a non-convex solver to produce an optimized output power vector for given input channels. Note that the approach does not directly optimize the performance criterion of interest, such as the sum-rate. Rather, it relies on the assumption that similar inputs -- the channel coefficients -- generally yield similar optimal solutions -- the power allocation vector. if the analytical model available based on domain knowledge is only a coarse approximation of the physical model, the resulting training set can be used to augment the data in order to carry out a preliminary training of a machine learning model \cite{2018arXiv180801672Z}\footnote{This can be thought of as an example of experience learning as part of small-sample learning techniques \cite{2018arXiv180804572S}.}.

For an application at a full-duplex transceiver, we refer to \cite{balatsoukas2017non}, which learns how to cancel self-interference in order to overcome the lack of well-established models for the transmitter-receiver chain of non-linearities.

\subsubsection{Link and Medium Access Control Layers}

At the medium access control layer, we highlight some applications of machine learning that tackle the lack of mathematical models for complex access protocols and communication environments. In \cite{2018arXiv180710495S}, a mechanism is proposed to predict whether a channel decoder will succeed on the basis of the outputs of the first few iterations of the iterative decoding process. This binary predictor is useful in order to request an early retransmission at the link layer using Automatic Retransmission Request (ARQ) or Hybrid ARQ (HARQ) in order to reduce latency. At the medium access control layer, data-aided methods can instead be used to predict the availability of spectrum in the presence of interfering incumbent devices with complex activation patterns for cognitive radio applications \cite{tumuluru2010neural} (see also \cite{del2016estimating}). An approach that leverages depth images to detect the availability of mmwave channels is proposed in \cite{okamoto2018machine}.

\subsubsection{Network and Application Layers}

A task that is particularly well-suited for machine learning is the caching of popular contents for reduced latency and network congestion \cite{chen2017echo}. Caching may take place at the edge and, more traditionally, within the core network segment. Caching at the edge has the advantage of catering directly to the preference of the local population of users, but it generally suffers from a reduced hit rate due to the smaller available storage capacity. Optimizing the selection of contents to be stored at the edge can be formulated as a classification problem that can benefit from a data-driven approach in order to adapt to the specific features of the local traffic \cite{chen2017echo}.

\subsection{At the Cloud}
We now turn to some relevant tasks to be carried out at the cloud at both network and application layers.
\subsubsection{Network}
The main task of the network layer is routing (see \cite{zorzi2015cognition} for further discussion). Considering a software-defined networking implementation, routing requires the availability at a network controller of information regarding the quality of individual communication links in the core network, as well as regarding the status of the queues at the network routers. In the presence of wireless or optical communications, the quality of a link may not be available at the network controller, but it may be predicted using available historical data \cite{wang2006supervised, musumeci2018survey} in the absence of agreed-upon dynamic availability models. In a similar manner, predicting congestion can be framed as a data-aided classification problem \cite{tang2018removing}.

\subsubsection{Application}
Finally, a relevant supervised learning task is that of traffic classification, whereby data streams are classified on the basis of some extracted features, such as packet sizes and inter-arrival times, in terms of their applications, e.g., Voice over IP. \cite{nguyen2008survey}

\section{Unsupervised Learning}\label{sec:unsup}

As introduced in Sec. \ref{sec:introduction}, unlike supervised learning, unsupervised learning tasks operate over unlabelled data sets consisting solely of the inputs $x_n$, with $n=1,...,N$, and the general goal is that of discovering properties of the data. We start this section by reviewing some of the typical specific unsupervised learning tasks. We then cover methodology, models, and learning, including advanced methods such as Generative Adversarial Networks (GANs) \cite{goodfellow2016deep}.

\subsection{Goals and Definitions}\label{sec:goalsunsup}
In unsupervised learning, taking a frequentist formulation (see Sec. \ref{sec:goals}), we are given a training set $\mathcal{D}$ consisting of $N$ i.i.d. samples $\mathrm{x}_{n}\sim p(x)$ with $n=1,...,N$ generated from an unknown true distribution $p(x)$. The high-level goal is that of learning some useful properties of the distribution $p(x)$. More specifically, we can identify the following tasks.

\begin{itemize}
\item \emph{Density estimation}: Density estimation aims at estimating directly the distribution $p(x)$. This may be useful, for example, for use in plug-in estimators of information-theoretic quantities, for the design of compression algorithms, or to detect outliers; 
\item\emph{Clustering}: Clustering aims at partitioning all points in the data set $\mathcal{D}$ in groups of similar objects, where the notion of similarity is domain-dependent;
\item\emph{Dimensionality reduction}, \emph{representation}, \emph{and feature extraction}: These three related tasks represent each data point $x_{n}$ in a different space, typically of lower dimensionality, in order to highlight independent explanatory factors and/or to ease visualization, interpretation, or the implementation of successive tasks, e.g., classification;
\item\emph{Generation of new samples}: Given the data set $\mathcal{D}$, we wish to learn a machine that produces samples that are approximately distributed according to $p(x)$. As an example, if the data set contains images of celebrities, the idea is to produce plausible images of non-existent celebrities. This can be useful, e.g., to produce artificial scenes for video parameterizes or films.
\end{itemize}

As suggested by the variety of tasks listed above, unsupervised learning does not have a formal unified formulation as supervised learning. Nevertheless, the general methodology follows three main steps in a manner similar to supervised learning (see Sec. \ref{sec:notknown}). In Step 1 (model selection), a model, or a hypothesis class, is selected, defining the inductive bias of the learning process. This is done by positing a family of probability distributions $p(x|\theta)$ parameterized by a vector $\theta$. In Step 2 (learning), the data $\mathcal{D}$ is used to optimize a learning criterion with the aim of choosing a value for the parameter vector $\theta$. Finally, in Step 3, the trained model is leveraged in order to carry out the task of interest, e.g., clustering or sample generation.

In the following, we discuss Step 1 (model selection) and Step 2 (learning). For the formulation of specific tasks to be carried out at Step 3, we refer to, e.g., \cite{Bishop,goodfellow2016deep,simeone2017brief}.

\subsection{Models}\label{sec:unsupmodel}

\begin{figure} 
 	\centering
 	\includegraphics[width=8.3cm]{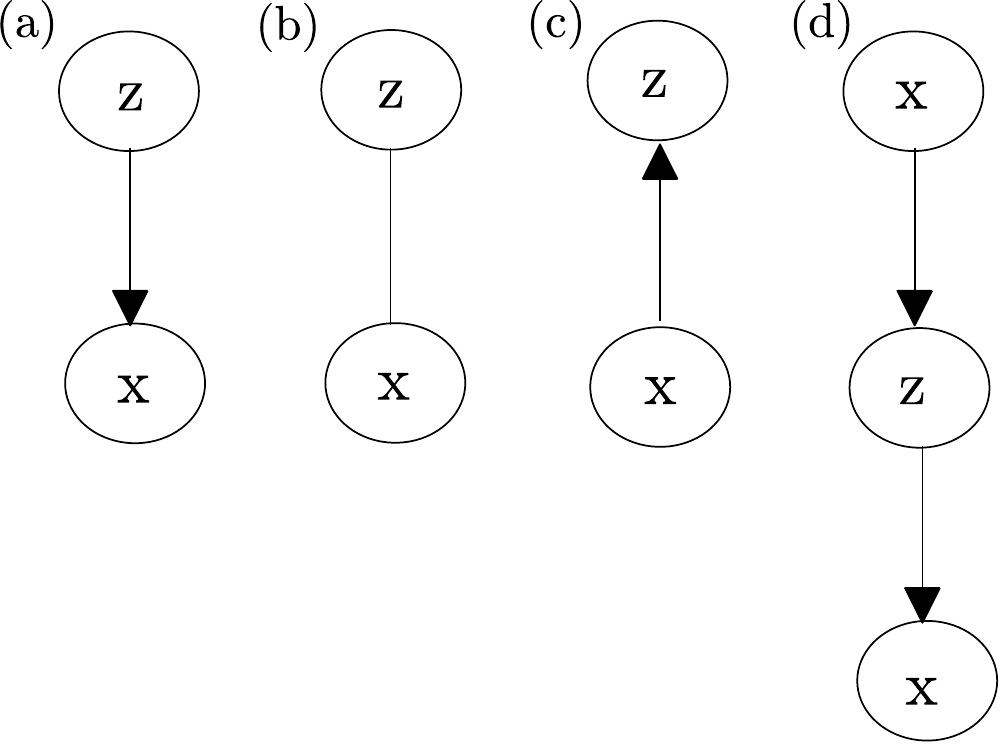}
 	\caption{Illustration of typical unsupervised learning models: (a) directed generative models; (b) undirected generative models; (c) discriminative models; and (d) autoencoders.}
 	\label{FigmodelsCh6}
 \end{figure}

Unsupervised learning models, selected at Step 1 of the machine learning process, typically involve a \emph{hidden or latent} (vector of) variables $\mathrm{z}_{n}$ for each data point $\mathrm{x}_n$. For example, in a clustering problem, the latent variable $\mathrm{z}_n$ represents the cluster index of $\mathrm{x}_n$. Latent variables are hidden or unobserved in the sense that they do not appear for any of the data points $x_n$ in $\mathcal{D}$.\footnote{Problems in which some of the inputs in $\mathcal{D}$ are labelled by a value $z_n$ are filed under the rubric of semi-supervised learning \cite{seeger2006taxonomy}.} The relationship between latent variables $\mathrm{z}_n$ and observable variables $\mathrm{x}_n$ can be modelled in different ways, giving rise to a number of different types of models for unsupervised learning. These are illustrated in Fig. \ref{FigmodelsCh6} and discussed next.
 
By way of a short round-up of types of models, with reference to Fig. \ref{FigmodelsCh6}, \emph{directed generative models}, illustrated by Fig. \ref{FigmodelsCh6}(a), posit that there exist hidden causes $\mathrm{z}$ yielding the observation $\mathrm{x}$. \emph{Undirected generative models}, represented in Fig. \ref{FigmodelsCh6}(b) model the mutual correlation between $\mathrm{x}$ and $\mathrm{z}$. \emph{Discriminative models}, illustrated by  Fig. \ref{FigmodelsCh6}(c) model the extraction of the latent representation $\mathrm{z}$ from $\mathrm{x}$. Finally, \emph{autoencoders}, represented in Fig. \ref{FigmodelsCh6}(d) assume that $\mathrm{x}$ is encoded into a latent representation $\mathrm{z}$ in such as way that $\mathrm{x}$ can then be approximately recovered from $\mathrm{z}$. In the following, we provide some additional details about directed generative models and autoencoders, and we point to \cite{simeone2017brief} and references therein for a discussion about the remaining models.

As illustrated in Fig. \ref{FigmodelsCh6}(a), directed generative models assume that each data point $\mathrm{x}$ is caused\footnote{The use of the term ``cause" is meant to be taken in an intuitive, rather than formal, way. For a discussion on the study of causality, we refer to \cite{pearl2018book}.} by a hidden variable $\mathrm{z}$. This is in the sense that the joint distribution $p(x,z|\theta)$ is parameterized as $p(x,z|\theta)=p(z|\theta)p(x|z,\theta)$, where $p(z|\theta)$ is the distribution of the hidden cause and $p(x|z,\theta)$ is the conditional distribution of the data $\mathrm{x}$ given the cause $\mathrm{z}$. As a result, under a directed generative model, the distribution of an observation $\mathrm{x}=x$ can be written as \begin{equation}\label{eq:dirgenmod}
p(x|\theta)=\sum_{z}p(z|\theta)p(x|z,\theta)=\textrm{E}_{\mathrm{z}\sim p(z|\theta)}[\ln p(x|\mathrm{z},\theta)],
\end{equation} where the sum in the second term should be replaced by an integration for continuous hidden variables, and the last equality expresses the marginalization over $\mathrm{z}$ as an expectation.

As an example, for the problem of document clustering, variable $\mathrm{x}$ represents a document in the training set and $\mathrm{z}$ is interpreted as a latent topic that ``causes" the generation of the document. Model selection requires the specification of a parameterized distribution $p(z|\theta)$ over the topics, e.g., a categorical distribution with parameters equals to the probability of each possible value, and the distribution $p(x|z,\theta)$ of the document given a topic. Basic representatives of directed generative models include mixture of Gaussians and likelihood-free models \cite{Murphy,simeone2017brief}.

As represented in Fig. \ref{FigmodelsCh6}(d), autoencoders model encoding from data $\mathrm{x}$ to hidden variables $\mathrm{z}$, as well as decoding from hidden variables back to data. Accordingly, model selection for autoencoders requires the specification of a parameterized family of encoders $p(z|x,\theta)$ and decoders $p(x|z,\theta)$. As an example, autoencoders can be used to learn how to compress an input signal $\mathrm{x}$ into a representation $\mathrm{z}$ in a smaller space so as to ensure that $\mathrm{x}$ can be recovered from $\mathrm{z}$ within an admissible level of distortion. Representatives of autoencoders, which correspond to specific choices for the encoder and decoder families of distributions, include Principal Component Analysis (PCA), dictionary learning, and neural network-based autoencoders \cite{Murphy,Bishop,simeone2017brief}.

\subsection{Learning}\label{sec:learnunsup}
We now discuss learning, to be carried out as Step 2. For brevity, we focus on directed generative models and refer to \cite{simeone2017brief} and references therein for a treatment of learning for the other models in Fig. \ref{FigmodelsCh6}. In this regard, we note that the problem of training autoencoders is akin to supervised learning in the sense that autoencoders specify the desired output for each input in the training set. 


As for supervised learning, the most basic learning criterion for probabilistic models is ML. Following the discussion in Sec. \ref{sec:learning}, ML tackles the problem of maximizing the log-likelihood of the data, i.e.,\begin{equation}\label{eq:MLunsup}
\underset{\theta}{\text{maximize }}\ln p(x|\theta)=\ln\textrm{E}_{\mathrm{z}\sim p(z|\theta)}[\ln p(x|\mathrm{z},\theta)].\end{equation}
Note that problem (\ref{eq:MLunsup}) considers only one data point $x$ in the data set for the purpose of simplifying the notation, but in practice the log-likelihood needs to be summed over the $N$ examples in $\mathcal{D}$. 

Unlike the corresponding problem for supervised learning (\ref{eq:MLprob}), the likelihood in (\ref{eq:MLunsup}) requires an average over the hidden variables. This is because the value of the hidden variables $\mathrm{z}$ is not known, and hence the probability of the observation $x$ needs to account for all possible values of $\mathrm{z}$ weighted by their probabilities $p(z|\theta)$. This creates a number of technical challenges. First, the objective in (\ref{eq:MLunsup}) is generally more complex to optimize, since the average over $\mathrm{z}$ destroys the typical structure of the model $p(x|z,\theta)$, whose logarithm is often selected as a tractable function (see, e.g., logistic regression). Second, the average in (\ref{eq:MLunsup}) cannot be directly approximated using Monte Carlo methods if the goal is to optimize  over the model parameters $\theta$, given that the distribution $p(z|\theta)$ generally depends on $\theta$ itself.

To tackle these issues, a standard approach is based on the introduction of a \emph{variational distribution} $q(z)$ over the hidden variables and on the optimization of a tractable lower bound on the log-likelihood known as the \emph{Evidence Lower BOund (ELBO)}. To elaborate, for any fixed value $x$ and any distribution $q(z)$ on the latent variables $\mathrm{z}$ (possibly dependent on $x$), the ELBO $\mathcal{L\mathrm{(\text{\ensuremath{q}},\theta)}}$
is defined as
\begin{equation}\label{eq:elbo}
\mathcal{L}(q,\theta)=  \textrm{E}_{\mathrm{z}\sim q(z)}[\ln p(x|\mathrm{z},\theta)]-\mathrm{KL}(q(\mathrm{z})||p(z|\theta)),
\end{equation} where $\mathrm{KL}(q||p)=\textrm{E}_{\mathrm{z}\sim q(z)}[\ln(q(\mathrm{z})/p(\mathrm{z}))]$ is the Kullback-Leibler (KL) divergence. The latter is a measure of the distance between the two distributions, as we will further discuss in Sec. \ref{sec:advanced}  (see \cite{cover2012elements,simeone2018introducing}). The analytical advantages of the ELBO $\mathcal{L}(q,\theta)$ over the original log-likelihood are that: (\emph{i}) it entails an expectation of the logarithm of the model $p(x|z,\theta)$, which, as mentioned, is typically a tractable function; and (\emph{ii}) the average is over a fixed distribution $q(z)$, which does not depend on the model parameter $\theta$. 

\begin{figure} 
	\centering
	\includegraphics[width=8.3cm]{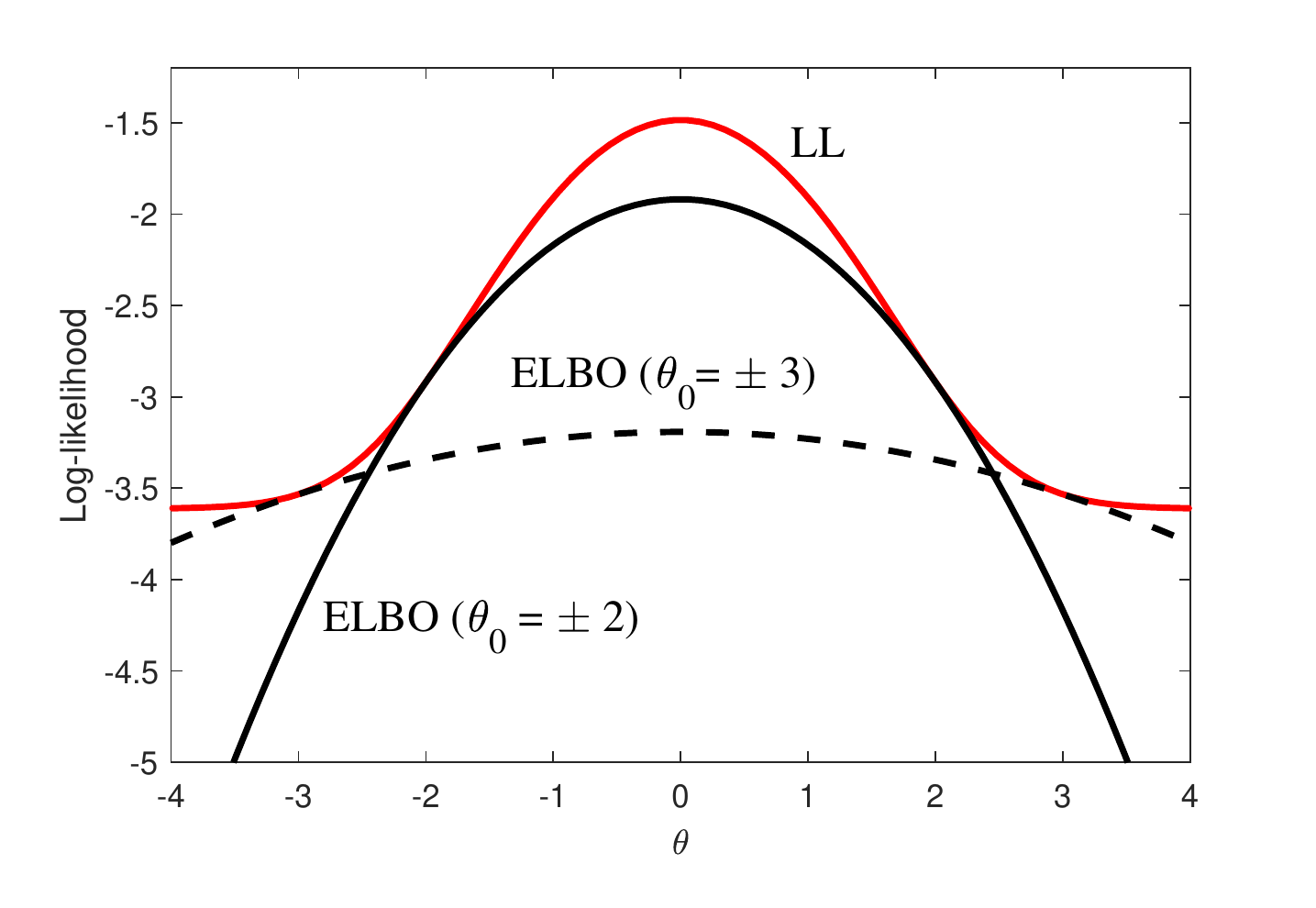}
	\caption{The ELBO (\ref{eq:elbo}) is a global lower bound on the log-likelihood that is tight at values of the model parameters $\theta_0$ for which equality (\ref{eq:eqelbo}) holds.}
	\label{FigELBOCh6}
\end{figure}

Using Jensen's inequality, it can be seen that the ELBO (\ref{eq:elbo}) is a global lower bound on the log-likelihood function, that is,
\begin{equation}\label{eq:lbelbo}
\text{\ensuremath{\ln p(x|}\ensuremath{\theta}) }\geq\mathcal{L}(q,\theta).
\end{equation}
An illustration of the lower bounding property of the ELBO can be found in Fig. \ref{FigELBOCh6}. An important feature of this inequality is that the ELBO ``touches" the log-likelihood function at values $\theta_{0}$, if any, for which the distribution $q(z)$ satisfies the equality
\begin{equation}\label{eq:eqelbo}
q(z)=p(z|x,\theta_{0}).
\end{equation} In words, the ELBO is tight if the variational distribution is selected to equal the posterior distribution of the hidden variables given the observation $x$ under the model parameter $\theta_0$. Stated less formally, in order to ensure that the ELBO is tight at a value $\theta_0$, one needs to solve the problem of inferring the distribution of the hidden variables $\mathrm{z}$ given the observation $x$ under the model identified by the value $\theta_0$.   

\begin{figure} 
	\centering
	\includegraphics[width=8.3cm]{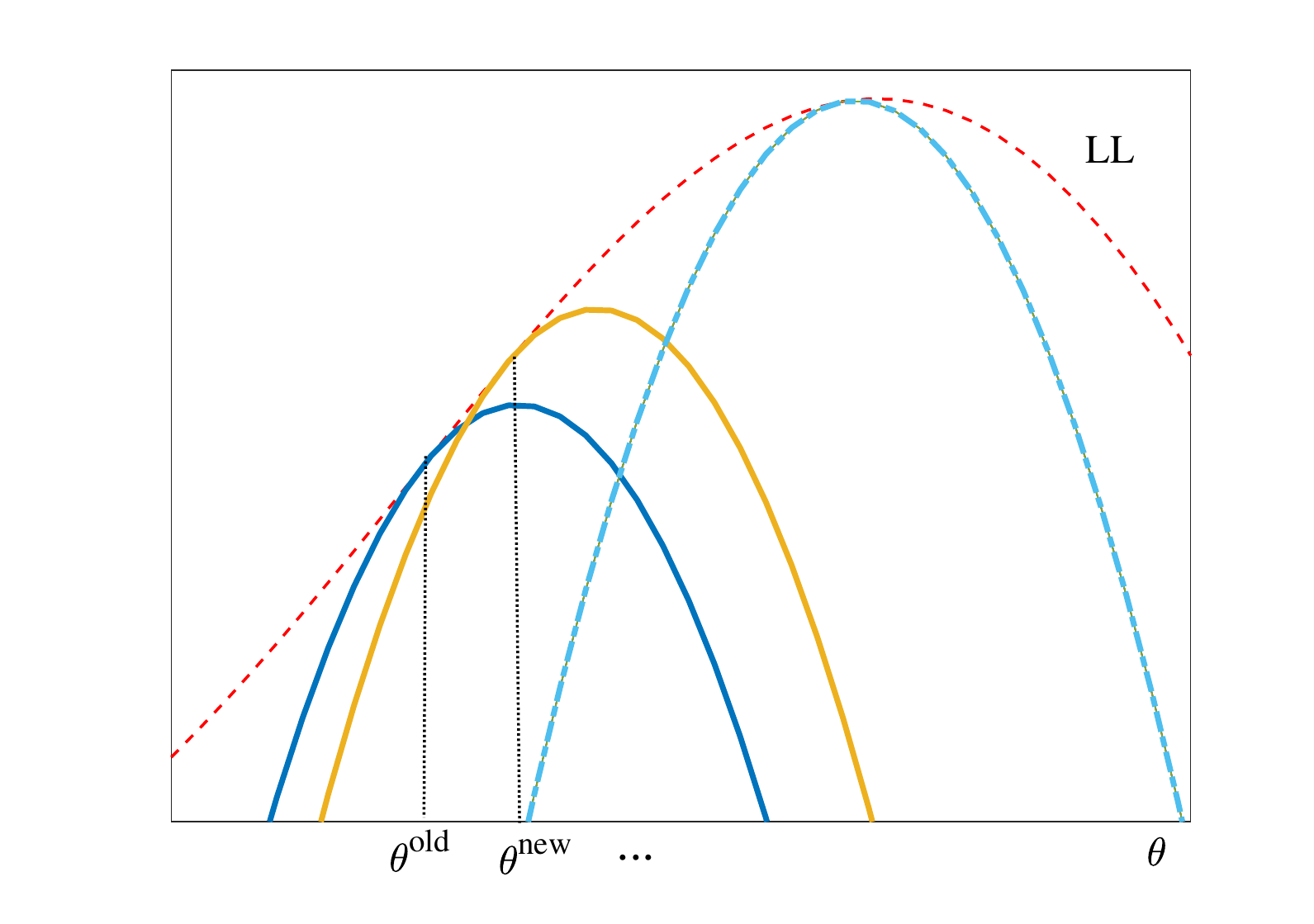}
	\caption{Illustration of the EM algorithm: At each iteration, a tight ELBO is evaluated in the E step by solving the problem of estimating the latent variables (via the posterior distribution $p(z|x,\theta)$), and then the ELBO is maximized in the M step by solving a problem akin to supervised learning with the estimated latent variables.}
	\label{FigMMCh6}
\end{figure}

The property (\ref{eq:lbelbo}) leads to the natural idea of the Expectation-Maximization (EM) algorithm as a means to tackle the ML problem. As illustrated in Fig. \ref{FigMMCh6}, EM maximizes the ELBO iteratively, where the ELBO at each iteration is computed to be tight at the current iterate for $\theta$. More formally, the EM algorithm can be summarized as follows\footnote{EM is an instance of the more general Majorization-Minimization algorithm \cite{sun2017majorization}.}. The model vector is initialized to some value $\theta^{\textrm{old}}$ and then for each iteration the following two steps are performed.
\begin{itemize}
	\item \emph{Expectation, or E, step:} For fixed parameter vector $\theta^{\textrm{old}}$, solve the problem \begin{equation}\label{eq:estep}
	\underset{q}{\text{maximize }}\mathcal{L}(q,\theta^{\textrm{old}}).
\end{equation}
The solution of this problem is given by $q^{\textrm{new}}(z)=p(z|x,\theta^{\textrm{old}})$. In fact, as discussed, the tightest (i.e., largest) value of the ELBO is obtained by choosing the variational distribution $q(z)$ as the posterior of the latent variables under the current model $\theta^{\mathrm{old}}$. This step can be interpreted as estimating the latent variables $\mathrm{z}$, via the predictive distribution $p(z|x,\theta^{\textrm{old}})$, assuming that the current model $\theta^{\textrm{old}}$ is correct.
	\item \emph{Maximization, or M, step:} For fixed variational distribution $q^{\textrm{new}}(z)$, solve the problem
	\begin{equation}\label{eq:mstep}
	\underset{\theta}{\text{maximize }}\mathcal{L}(q^{\textrm{new}},\theta)= \textrm{E}_{\mathrm{z}\sim q^{\textrm{new}}(z)}\left[\ln p(x,\mathrm{z}|\theta)\right].
	\end{equation} This optimization is akin to that carried out in the corresponding supervised learning problem with known latent variables $\mathrm{z}$ with the difference that these are randomly selected from the fixed variational distribution $q^{\textrm{new}}(z)$ obtained in the E step.
\end{itemize}
 
 Given that the EM algorithm maximizes at each step a lower bound on the log-likelihood that is tight at the current iterate $\theta^{\textrm{old}}$, EM guarantees decreasing objective values along the iterations, which ensures convergence to a local optimum of the original problem. We refer to \cite{Bishop,Murphy} for detailed examples.

The EM algorithm is generally  impractical for large-scale problems due to the complexity of computing the posterior of the latent variables in the E step and of averaging over such distribution in the M step. Many state-of-the-art solutions to the problem of unsupervised learning with probabilistic models entail some approximation of the EM algorithm. Notably, the E step can be approximated by parametrizing the variational distribution with some function $q(z|\varphi)$, or $q(z|x,\varphi)$ to include the dependence on $x$, and by maximizing ELBO over the variational parameters $\varphi$. This approach underlies the popular variational autoencoder technique \cite{goodfellow2016deep}. In the M step, instead, one can approximate the expectation in (\ref{eq:mstep}) using Monte Carlo stochastic approximation based on randomly sampled values of $\mathrm{z}$ from the current distribution $q(z)$. Finally, gradient descent can be used to carry out the mentioned optimizations for both E and M steps (see, e.g., \cite{mnih2014neural}).
 
\subsection{Advanced Learning Methods} \label{sec:advanced}
\begin{figure} 
	\centering
	\includegraphics[width=8.3cm]{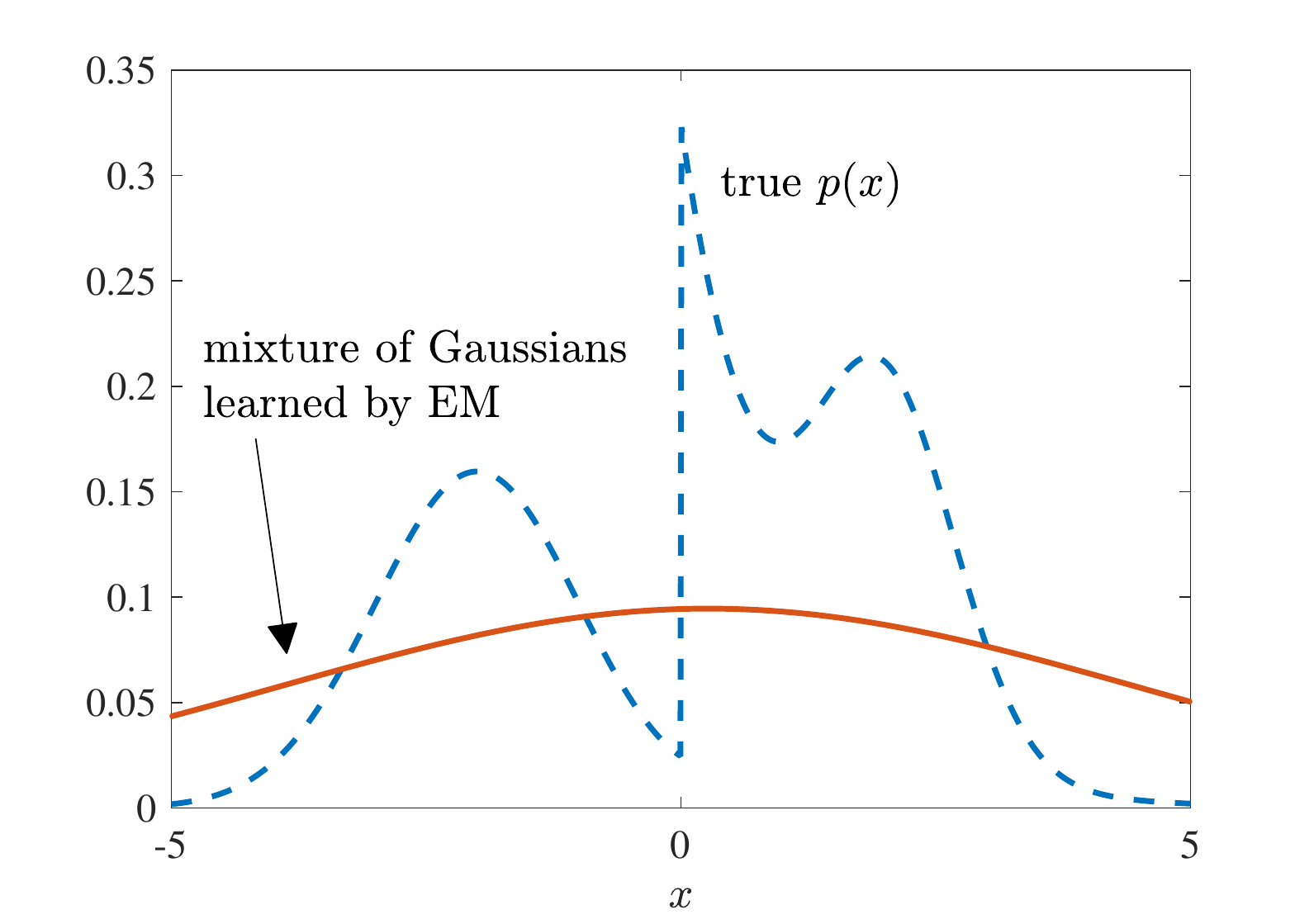}
	\caption{Illustration of the limitations of ML unsupervised learning, here obtained via the EM algorithm: The ML solution tends to be blurry, missing the modes of the true distribution $p(x)$.}
	\label{figmixGaussCh6}
\end{figure}

As discussed in the previous section, ML is generally prone to overfitting for supervised learning. For unsupervised learning, the performance of ML depends on the task of interest. For example, consider the tasks of density estimation or of generation of new samples (see Sec. \ref{sec:goalsunsup}). In order to illustrate some of the typical issues encountered when applying the ML criterion, in Fig. \ref{figmixGaussCh6} we report a numerical result for a problem in which the true data distribution $p(x)$ is multi-modal and the model distribution $p(x|\theta)$ is assumed to be a mixture of Gaussians, i.e., a directed generative model. The ML problem is tackled by using EM based on samples generated from the true distribution (see \cite{simeone2017brief} for details). The learned distribution is seen to be a rather``blurry" estimate that misses the modes of $p(x)$ in an attempt of being inclusive of the full support of $p(x)$. Being a poor estimate of the true distribution, the learned model can clearly also be problematic for sample generation in the sense that samples generated from the model would tend to be quite different from the data samples. In the rest of this section, we briefly review advanced learning methods that address this limitation of ML.

In order to move beyond ML, we first observe that ML can be proven to minimize the KL divergence \begin{equation}\label{eq:KLdiv}
\textrm{KL}(p_{\mathcal{D}}(x)||p(x|\theta))=\textrm{E}_{\mathrm{z}\sim p_{\mathcal{D}}(x)}\left[\ln\frac{p_{\mathcal{D}}(x)}{p(x|\theta)}\right]
\end{equation}
between the empirical distribution, or histogram, of the data\begin{equation}
p_{\mathcal{D}}(x)=\frac{N[x]}{N},
\end{equation}
where $N[x]$ counts the number of occurrences of value $x$ in the data, and the parameterized model distribution $p(x|\theta)$. In other words, ML fits the model to the histogram of the data by using the KL divergence as a measure of fitness. Indeed, as mentioned in Sec. \ref{sec:learnunsup}, the KL divergence is a quantitative measure of ``difference" between two distributions. More precisely, as per (\ref{eq:KLdiv}), the KL divergence $\textrm{KL}(p||q)$ quantifies the difference between two distributions $p(x)$ and $q(x)$ by evaluating the average of the LLR $\ln(p(x)/q(x))$ with respect to $p(x)$. 

\begin{figure} 
	\centering
	\includegraphics[width=8.3cm]{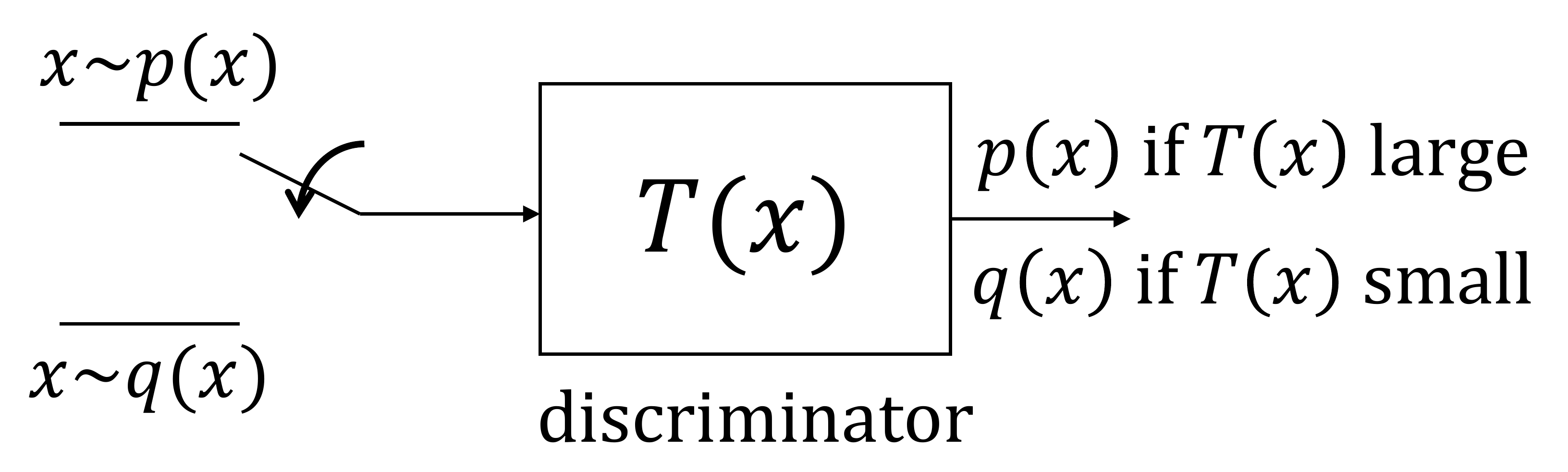}
	\caption{Discriminator between the hypotheses $\mathrm{x}\sim p(x)$ and $\mathrm{x}\sim q(x)$ based on the statistic $T(x)$. The performance of the optimal discriminator function $T(x)$ under different design criteria yields a measure of the difference between the two distributions. }
	\label{discriminator}
\end{figure}

Consider now the problem illustrated in Fig. \ref{discriminator}, in which a discriminator wishes to distinguish between two hypotheses, namely the hypothesis that the data $x$ is a sample from distribution $p(x)$ and the hypothesis that it is instead generated from $q(x)$. To fix the ideas, one can focus as an example on the case where $p(x)$ and $q(x)$ are two Gaussian distributions with different means. To this end, the discriminator computes a statistic, that is, a function, $T(x)$ of the data $x$, and then decides for the former hypothesis if $T(x)$ is sufficiently large and for the latter hypothesis otherwise. Intuitively, one should expect that, the more distinct the two distributions $p(x)$ and $q(x)$ are, the easier it is to design a discriminator that is able to choose the correct hypothesis with high probability.

The connection between the hypothesis testing problem in Fig. \ref{discriminator} and the KL divergence becomes evident if one recalls that the LLR $\ln(p(x)/q(x))$ is known to be the best statistic $T(x)$ in the Neyman-Pearson sense \cite{poor2013introduction}. The KL divergence is hence associated to a particular way of evaluating the performance of the discriminator between the two distributions. Considering a broader formulation of the problem of designing the discriminator in Fig. \ref{discriminator}, one can generalize the notion of KL divergence to the class of $f$-divergences. These are defined as \begin{equation}\label{eq:fdivergence}
D_{f}(p||q)=\max_{T(x)}\textrm{E}_{\mathrm{x}\sim p(x)}[T(\mathrm{x})]-\textrm{E}_{\mathrm{x}\sim q(x)}[g(T(\mathrm{x}))],
\end{equation}
for some concave increasing function $g(\cdot)$. The expression above can be interpreted as measuring the performance of the best discriminator $T(x)$ when the design criterion is given by the right-hand side of (\ref{eq:fdivergence}), i.e., $\textrm{E}_{\mathrm{x}\sim p(x)}[T(\mathrm{x})]-\textrm{E}_{\mathrm{x}\sim q(x)}[g(T(\mathrm{x}))]$, for a given function $g(\cdot)$. Note that this criterion is indeed larger for a discriminator that is able to output a large value of the statistic $T(x)$ under $p(x)$ and a small value under $q(x)$. The KL divergence corresponds to a specific choice of such function (see \cite{simeone2017brief} for details).

In order to move beyond ML, one can then consider fitting the model distribution to the data histogram by using a divergence measure that is tailored to the data and that captures the features of the empirical distribution that are most relevant for a given application. Such a divergence measure can be obtained by choosing a suitable function  $g(\cdot)$ in (\ref{eq:fdivergence}) and by optimizing (\ref{eq:fdivergence}) over a parameterized (differentiable) discriminator function $T_{\varphi}(\mathrm{x})$. Integrating the evaluation of the divergence with the problem of learning the model parameters yields the min-max problem \begin{equation}
\textrm{\ensuremath{\underset{\theta}{\min}} }\max_{\varphi}\textrm{E}_{\mathrm{x}\sim p_{\mathcal{D}}(x)}[T_{\varphi}(\mathrm{x})]-\textrm{E}_{\mathrm{x}\sim p(x|\theta)}[g(T_{\varphi}(\mathrm{x}))].
\end{equation} This can be famously interpreted as a game between the learner, which optimizes the model parameters $\theta$, and the discriminator, which tries to find the best function $T_{\varphi}(\mathrm{x})$ to distinguish between data and generated samples. The resulting method, known as GAN, has recently led to impressive improvements of ML for sample generation \cite{goodfellow2016nips}.

\section{Applications of Unsupervised Learning to Communication Systems}\label{sec:unsupappl}
In this section, we highlight some applications of unsupervised learning to communication networks. 

\subsection{At the Edge}

\subsubsection{Physical Layer} 
Let us first consider some applications of \emph{autoencoders} at the physical layer as implemented by the network edge nodes. A fundamental idea is to treat the chain of encoder, channel, and decoder in a communication link as an autoencoder, where, with reference to Fig. \ref{FigmodelsCh6}(d), the input message is $x$, the transmitted codewords and received signals represent the intermediate representation $z$, and the output of the decoder should match the input \cite{o2017introduction}. Note that, for this particular autoencoder, the mapping $p(x|z)$ can only be partially learned, as it includes not only the encoder but also the communication channel, while the conditional distribution $p(x|z)$ defining the decoder can be learned. We should now ask when this viewpoint can be beneficial in light of the criteria reviewed in Sec. \ref{sec:when}.

To address this question, one should check whether a model or algorithm deficit exists to justify the use of machine learning tools. Training an autoencoder requires the availability of a model for the channel, and hence a model deficit would make this approach inapplicable unless further mechanisms are put in place (see below). Examples of algorithm deficit include channels with complex non-linear dynamical models, such as optical links \cite{karanov2018end}; Gaussian channels with feedback, for which optimal practical encoding schemes are not known \cite{kim2018deepcode}; multiple access channels with sparse transmission codes \cite{scma}; and joint source-channel coding \cite{2018arXiv180901733B}.

Other applications at the physical layer leverage the use of autoencoders as compressors (see Sec. \ref{sec:unsupmodel}) or denoisers. For channels with a complex structure with unavailable channel models or with unknown optimal compression algorithms, autoencoders can be used to compress channel state information for the purpose of feedback on frequency-division duplex links \cite{wen2018deep}. Autoencoders can also be used for their capacity to denoise the input signal by means of filtering through the lower dimensional representation $z$. This is done in \cite{xiao20173} for the task of localization on the basis of the received baseband signal. To this end, an autoencoder is learned for every reference position in space with the objective of denoising signals received from the given location. At test time, the location that corresponds to the autoencoder with the smallest reconstruction error is taken as an estimate of the unknown transmitting device.

We now review some applications of the \emph{generative models} illustrated in Fig. \ref{FigmodelsCh6}(a). A natural idea is that of using generative models to learn how to generate samples from a given channel \cite{o2018approximating,ye2018channel}. This approach is sound for scenarios that lack tractable channel models. As a pertinent example, generative models can be used to mimic and identify non-linear channels for satellite communications \cite{ibnkahla2000applications}. The early works on the subject carried out in the nineties are also notable for the integration of the domain knowledge into the definition of machine learning models (see Sec. \ref{sec:supappl}). In fact, mindful of the strong linear components of the channels, these works posit a learnable model that includes linear filters and non-linearities \cite{ibnkahla2000applications}.

Another approach that can be considered as unsupervised was proposed in \cite{2018arXiv180710025L} in order to solve the challenging problem of power control for interference channels. The approach tackles the resulting algorithm deficit by means of a direct optimization of the sum-rate with the aim of obtaining the power allocation vector (as fractions of the maximal available powers) at the output of a neural network. Related supervised learning methods were discussed in Sec. \ref{sec:supappl}. A similar approach -- also based on the idea of directly maximizing the criterion of interest so as to obtain an approximate solution at the output of a neural network -- was considered in \cite{neumann2018learning} for minimum mean squared error channel estimation with non-Gaussian channels, e.g., multi-path channels. 

\subsubsection{Medium Access Layer}
At the medium access layer, generative models have been advocated in  \cite{davaslioglu2018generative} as a way to generate new examples so as to augment a data set used to train a classifier for spectrum sensing (see Sec. \ref{sec:supappl}). An unsupervised learning task that has found many applications in communications is \emph{clustering}. For example, in \cite{abdelnasser2014clustering}, clustering is used to support radio resource allocation in a heterogeneous network.

\subsection{At the Cloud}

\subsubsection{Network Layer}
Another typical application of clustering is to enable hierarchical clustering for routing in self-organizing multi-hop networks. Thanks to clustering, routing can be carried out more efficiently by routing first at the level of clusters, and then locally within each cluster \cite{abbasi2007survey}. For an application of the unsupervised learning task of \emph{density estimation}, consider the problem of detecting anomalies in networks. For instance, by learning the typical distribution of the features of a working link, one can identify malfunctioning ones. This approach may be applied, e.g., to optical networks \cite{musumeci2018survey}.

\subsubsection{Application Layer}
Finally, we point to two instances of unsupervised learning at the application layer that are usually carried out at data centers in the cloud. These tasks follow a conceptually different approach as they are based on discovering structure in graphs. The first problem is community detection in social networks. This amounts to a clustering problem whereby one wishes to isolate communities of nodes in a social graph on the basis of the observation of a realization of the underlying true graph of relationships \cite{abbe2014exact}. Another application is the ranking of webpages based on the graph of hyperlinks carried out by PageRank \cite{page1999pagerank,simeone2017brief}.

\section{Concluding Remarks}\label{sec:conclusions}

In the presence of modelling or algorithmic deficiencies in the conventional engineering flow based on the acquisition of domain knowledge, data-driven machine learning tools can speed up the design cycle, reduce the complexity and cost of implementation, and improve over the performance of known algorithms. To this end, machine learning can leverage the availability of data and computing resources in many engineering domains, including modern communication systems. Supervised, unsupervised, and reinforcement learning paradigms lend themselves to different tasks depending on the availability of examples of desired behaviour or of feedback. The applicability of learning methods hinges on specific features of the problem under study, including its time variability and its tolerance to errors. As such, a data-driven approach should not be considered as a universal solution, but rather as a useful tool whose suitability should be assessed on a case-by-case basis. Furthermore, machine learning tools allow for the integration of traditional model-based engineering techniques and of existing domain knowledge in order to leverage the complementarity and synergy of the two solutions (see Fig. \ref{FigMLvENGcon}).

As a final note, while this paper has focused on applications of machine learning to communication systems, communication is conversely a key element of distributed machine learning platforms. In these systems, learning tasks are carried out at distributed machines that need to coordinate via communication, e.g., by transferring the results of intermediate computations. A recent line of work investigates the resulting interplay between computation and communication \cite{diggavi}.

\bibliographystyle{IEEEtran}
\bibliography{references}

\end{document}